\documentclass[letterpaper,12pt]{article}

\usepackage{endnotes}
\usepackage{amssymb}
\usepackage{amsfonts}
\usepackage{amsmath}
\usepackage{amsthm}
\usepackage{graphicx,color}
\usepackage{xcolor}
\usepackage{comment} 
\usepackage{chicago}
\usepackage{array}

\usepackage{url}

\usepackage{setspace}
\def\R{\mathbb{R}}
\def\endproof{\hfill\diamondsuit}

\def\sF{{\mathcal F}}

\def\sA{{\mathcal A}}

\def\E{\mathbb{E}}
\def\V{\mathbb{V}}
\def\sF{\mathcal{F}}

\def\N{\mathbb N}

\newcommand{\ta}{\tilde{a}} 
\newcommand{\taS}{\tilde{a}_\Sigma}

\newcommand{\bM}{\bar{M}}
\newcommand{\bk}{\bar{\kappa}}
\newcommand{\bC}{\bar{C}}
\newcommand{\cM}{M^\circ}
\newcommand{\ck}{\kappa^\circ}
\newcommand{\ce}{\epsilon^\circ}
\newcommand{\cC}{C^\circ}

\numberwithin{equation}{section}
\theoremstyle{plain}                % title and  number in bold, text italic
\newtheorem{theorem}{Theorem}[section]
\newtheorem{lemma}[theorem]{Lemma}

\newtheorem{corollary}[theorem]{Corollary}
\theoremstyle{definition}           % title and number in bold, text normal
\newtheorem{definition}[theorem]{Definition}
\newtheorem{example}[theorem]{Example}

\theoremstyle{remark}               % title and number in italic, text normal
\newtheorem{remark}{Remark}[section]

\usepackage[margin=1.25in]{geometry}

\usepackage{ulem} 

\begin{document}
\pagestyle{empty}

\begin{center}
\large{\bf Equilibrium Effects of Intraday Order-Splitting Benchmarks}$^\ast$\makeatletter{\renewcommand*{\@makefnmark}{}\footnotetext{\hspace{-.35in} $^\ast${The authors have benefited from helpful comments from Yashar Barardehi, Umut Cetin, Chris Frei, Steve Karolyi, Ajitesh Mehta, Johannes Muhle-Karbe, Dan Ocone, Tom Ruchti, Chester Spatt, Harvey Stein, Kim Weston, Ariel Zetlin-Jones, and seminar participants at Carnegie Mellon University, Columbia University, IAQF/Thalesians, and Intech. Kasper Larsen is partly supported by the National Science Foundation (NSF) under Grant No. DMS 1812679 (2018 - 2021). Any opinions, findings, and conclusions or recommendations expressed in this material are those of the authors and do not necessarily reflect the views of the National Science Foundation (NSF). Jin Hyuk Choi has email: jchoi@unist.ac.kr and Duane J. Seppi has email: ds64@andrew.cmu.edu.  The corresponding author is Kasper Larsen.  Kasper Larsen has email: KL756@math.rutgers.edu.}\makeatother}}
\end{center}
\vspace{0.25cm}

\begin{center}

Jin Hyuk Choi\\ Ulsan National Institute of Science and Technology (UNIST)

\ \\

Kasper Larsen \\  Rutgers University

\ \\

Duane J. Seppi\\ Carnegie Mellon University

\end{center}

\begin{center}

 \today

\end{center}

\ \\

\begin{verse}

{\sc Abstract}: This paper presents a continuous-time model of  intraday trading, pricing, and liquidity with dynamic  TWAP and VWAP benchmarks. The model is solved in closed-form for the competitive equilibrium and also for non-price-taking equilibria. The intraday trajectories of TWAP trading targets cause predictable intraday patterns of price pressure, and randomness in VWAP target trajectories induces additional randomness in intraday price-pressure patterns.  TWAP and VWAP trading both reduce market liquidity and increase price volatility relative to just terminal trading targets alone. The model is computationally tractable, which lets us provide a number of numerical illustrations.

\end{verse}
\begin{verse}

%\vspace{0.2cm}

%{\sc AMS subject classifications}: 93E20

%\vspace{0.2cm}

%{\sc JEL-Classification}: G12, G11, D53
\end{verse}

%\end{document}
%\newpage
%Disclosure Statements:\ \\

%Jin Hyuk Choi: Nothing to disclose regarding outside support or conflicts of interest.\ \\

%Kasper Larsen: Is supported by the National Science Foundation (NSF) under Grant No. DMS 1812679 (2018 - 2021).\ \\

%Duane Seppi: Nothing to disclose regarding outside support or conflicts of interest. 

\newpage
\pagestyle{plain}
\addtocounter{page}{-1}

%\section{Introduction} 
Dynamic order-splitting strategies are a prominent feature of present-day financial markets. As described in O'Hara (2015), large asset managers use sequences of small {\em child orders} to trade on large latent meta {\em parent demands}. These strategies include heuristically mimicking the time-weighted average price (TWAP) or volume-weighted average price (VWAP) and also optimized strategies from formal execution-cost minimization problems (see, e.g., Almgren and Chriss, 1999 and 2000). The scale of order splitting in present-day markets is substantial. The Financial Insights (2006) trading survey reports that one popular strategy, VWAP execution orders, represents around 50\% of all institutional trading.\footnote{Large asset managers conduct dynamic trading strategies using in-house trading desks and also via principal and agency trading with external brokers.}   Dynamic trading strategies can be price-sensitive in that they respond to the price of liquidity as investors trade off price improvement and trading profits vs. tracking error  relative to intraday trading targets.\footnote{Madhavan (2002) discusses price improvement on order execution relative to VWAP. Domowitz and Yegerman (2005) estimate empirical order-execution costs benchmarked relative to VWAP.} In addition, high frequency market makers use dynamic  strategies to supply liquidity while attempting to keep their inventory close to an intraday target of zero.\footnote{Hagstr\"omer and Nord\'en  (2013) and Menkveld (2013) show that high-frequency (HFT) market makers are an important source of intraday liquidity.  A common feature of HFT market makers is that they have ``very short time-frames for establishing and liquidating positions" (SEC 2010), which is consistent with a zero target inventory level. Weller (2013) shows further that liquidity over the trading day is provided by a network of liquidity providers with slower and faster trading latencies who shift inventory between themselves over holding periods of different lengths. This behavior is also consistent with zero intraday inventory targets.}   The result is a complex ecosystem in which asset managers, other position-taking traders, and high frequency market makers all use dynamic trading strategies to supply and demand liquidity given different benchmarked intraday targets.

Our paper is the first to model the equilibrium impact of TWAP, VWAP, and other dynamic benchmarks on intraday trading and liquidity. Specifically, we focus on liquidity effects of order splitting rather than on information aggregation as in Kyle (1985).
We take intraday trading target trajectories and penalties as inputs to our model, and then show how they affect price dynamics and liquidity over the trading day via their effect on aggregate order imbalances. In particular, we model a market with multiple price-sensitive  investors with heterogeneous trading targets who follow optimal continuous-time dynamic strategies. We solve for equilibria in closed-form. Our main results are as follows:

\begin{itemize}

\item Order-splitting benchmarks like TWAP, Almgren-Chriss targets, and VWAP lead to predictable intraday price pressure due to persistant latent trading-demand imbalances.  In particular, aggregate latent buy (sell) imbalances lead to positive (negative) price pressure at the market open that then decays, i.e., trends down (up) on average, over the trading day.

\item  Trading benchmarks reduce intraday liquidity and increase price volatility relative to when investors just have terminal end-of-day trading targets. This is because penalties for intraday deviations from trading target trajectories reduce the inventory-holding flexibility of investors over the day.

\item Comparative static results lead to empirical predictions about price dynamics. For example, market illiquidity and price volatility in a competitive Radner equilibrium are increasing in investor order-execution costs and inventory penalties for trading deviations from trading target trajectories. 

\item In a competitive Radner equilibrium, price-sensitive investors deviate from their ideal trading targets by a fraction of the aggregate latent trading-demand imbalance. In non-price-taking equilibria, investor target deviations can also depend on individual investor targets.

\end{itemize} 

\noindent

Our research is related to several prior research literatures. First, there is a large optimal control literature on optimal order execution with exogenous price impact. This includes Bertsimas and Lo (1998); Almgren and Chriss (1999, 2000); Gatheral and Schied (2011); Engle et al., (2012); Predoiu et al., (2011); Boulatov et al., (2016) and other research surveyed in Gatheral and Shied (2013). In contrast, we model dynamic trading benchmarks and their effects in an equilibrium framework. Second, the practice of benchmarking order-execution quality with VWAP and other metrics is described in Berkowitz, Logue, and Noser (1988) and Madhavan (2002).\footnote{Implementation shortfall is another alternative trading benchmark; see Perold (1988).} Baldauf, Frei, and Mollner (2018) show that VWAP benchmarking is optimal for certain principal-agent problems in delegated order execution. In contrast, we model the market effects of trading benchmarks rather than the reasons why investors use such strategies. Third, Korajczyk and Murphy (2019), van Kervel and Menkveld (2018), and van Kervel, Kwan, and Westerholm (2018) document empirical interactions between dynamic trading by different investors. Our work is most closely related to van Kerval et al. (2018), which shows theoretically and empirically how dynamic trading strategies interact across multiple strategic investors.  In contrast, we model the equilibrium price effects of dynamic trading as well as trading interaction effects.

Our analysis extends the costly inventory model of market making --- see, e.g., Garman (1976), Stoll (1978), and Grossman and Miller (1988) --- by allowing for endogenous arriving orders from optimized order-splitting strategies. In particular, our model highlights the idea that, with endogenous investor trading demand, price pressure in equilibrium prices is not just due to inventory-holding costs of liquidity providers on executed transactions. Rather, equilibrium price pressure also depends on the underlying latent investor trading-demand imbalances. In order for markets to clear, equilibrium prices deter investors from trading on trading demands for which is there is no counterparty and, thus, which do not lead to executed transactions.  This is a new perspective relative to more transactional market microstructure models, but one that is consistent with empirical evidence in van Kerval et al. (2018) that large investors trade less in the direction of aggregate imbalances in underlying parent trading demands.    Our modeling approach is related to previous research by Brunnermeier and Pedersen (2005) and Carlin, Lobo, and Viswanathan (2007) on optimal rebalancing and predatory trading.  There are two main differences: First, investors in our model are subject to penalties tied to intraday trading target trajectories rather than to just a single hard terminal constraint at the end of the day. Second, there are no ad hoc intraday liquidity providers in our model. Instead, all intraday liquidity is provided endogenously by rational optimizing investors. As a result, there is no predatory trading in our model even when investors are strategic.\footnote{Predatory trading is a manipulative strategy to buy (sell) to raise (lower) prices artificially and then unwind these positions at a profit given price support from predictable persistent buying (selling) by a large investor rebalancing its portfolio. In Brunnermeier and Pedersen (2005), predatory trading is possible because of ad hoc liquidity providers who trade using exogenous linear schedules that do not rationally anticipate predictable future price changes later in the day given earlier trading.} Our paper also builds on the Vayanos (1999) dynamic trading model. Like Vayanos (1999) --- and the continuous-time extension in Sannikov and  Skrzypacz (2016) --- our model includes investors who smooth a series of random intraperiod shocks but, in addition, we also have investors with private heterogeneous trading target trajectories they would like to follow over the day. The discrete-time model in Du and Zhu (2017) also has quadratic penalties but with zero targets and has private dividend information (whereas dividend information is public in our model). 

Our model is also related to G\^arleanu and Pedersen (2016) and the multi-investor extension in Bouchard, Fukasawa, Herdegen, and Muhle-Karbe (2018). In these  competitive equilibrium models, traders incur penalties based on their stock holdings and their trading rates. Our model differs because we allow for possible price-impact of individual trades and for non-zero trading targets that are ex ante private information. 

Lastly, there is no asymmetric information about future asset cash-flow fundamentals in our model.  Thus, our analysis here on trading and non-informational price pressure is complementary to Choi, Larsen, and Seppi (2019), which studies order-splitting and dynamic rebalancing in a Kyle (1985) style market in which a strategic informed investor with long-lived private information and a strategic rebalancer with a hard terminal trading target both follow dynamic trading strategies.  

\section{Model primitives} \label{sec:1}
The goal of our analysis is to model the equilibrium effects of TWAP and other intraday trading benchmarks on prices and trading.  Thus, the necessary model ingredients are investors who have heterogeneous trading targets and penalties for deviating from their intraday targets and who are price-sensitive so that market-clearing can be endogenized. Moreover, given the high empirical frequency of child orders in dynamic order-execution strategies in Korajczyk and Murphy (2019) and van Kervel and Menkveld (2018), modeling these effects in continuous time is appropriate.

We develop a continuous-time equilibrium model with a unit time horizon in which trade takes place at each time point $t\in[0,1]$. This can be interpreted as one trading day. There are two securities: A money market account with a constant unit price (i.e., the account pays a zero interest rate) and a stock with an endogenously determined price process $S = (S_t)_{t\in[0,1]}$. Information about future dividends is generated by an exogenous standard Brownian motion $D=(D_t)_{t\in[0,1]}$ with a given known initial value $D_0 \in\R$, zero drift, and volatility normalized to one. Here $D_t$ denotes the expected present value at time $t \le 1$ of subsequent future dividends, where $D_1$ is the expectation at the end of the day. The market microstructure literature calls $D_t$ the ``fundamental asset value." In our model, the difference $S_t - D_t$ is price pressure required to clear the stock market.

Investors in our model all follow optimal trading strategies that trade off trading profits/costs and different types of inventory-holding penalties.  For a generic investor $i$, let $\theta_{i,t}$ denote the investor's actual stock holdings at time $t\in[0,1]$. By subtracting any initial stock position, we can, without loss of generality,  normalize initial stock positions $\theta_{i,-}$ to zero. Let $\theta^c_{i,-}$ denote the trader's initial cash money-market balance. Formally, the optimal stock holdings for each investor $i$ over the day solve an optimization problem
\begin{align}\label{objective}
V_i(X_{i,0}):=\sup_{\theta_{i}\in\sA_i}\E\left[X_{i,1} - L_{i,1}\,\Big|\,\sF_{i,0}\right],%\quad i\in\{1,...,M+\overline{M}\},
\end{align}
where $X_{i,1}$ is the investor's terminal wealth at $t=1$ given her trading gains or losses  over the day given by the wealth process 
\begin{align}\label{dX}
\begin{split}
X_{i,t} &:= \theta^c_{i,-} + \int_0^t \theta_{i,u} dS_u,\quad t\in[0,1],
\end{split}
\end{align}
where $X_{i,0} := \theta^c_{i,-}$ is investor $i$'s initial wealth (i.e., initial cash balance in her money market account) and where $L_{i,1}$ in \eqref{objective} is an investor-specific  terminal cumulative inventory penalty at $t=1$  that accrues over the day. The initial information set $\sF_{i,0}$ in \eqref{objective}  is formally defined in \eqref{meas1}, and the admissible set of controls $\sA_i$ is defined in Definition \ref{def_admis}.

Two types of investors trade in our model with different holding penalties:

\begin{itemize}

\item There are $M\in \N$ price-sensitive \emph{targeted investors} whose terminal penalty  $L_{i,1}$ accrues via a penalty process 
\begin{align}\label{defL}
L_{i,t}:=\int_0^t \kappa(s)\big(\theta_{i,s}-\gamma(s)\tilde{a}_i\big)^2ds,\quad t\in[0,1],\quad i\in\{1,...,M\},
\end{align} 
where $\ta_i$ is investor $i$'s end-of-day target stock holdings and $\gamma(t)\ta_i$ is a trajectory of intraday target stock holdings. The function $\kappa(t)$ describes the severity of penalties for  deviations of actual holdings $\theta_{i,t}$ over time from the target trajectory $\gamma(t)\ta_i$. The variables $(\ta_i,\theta^c_{i,-})$ are private knowledge of trader $i$ but are partially revealed in equilibrium to other investors. At this point, we make no distributional assumptions about $(\ta_1,...,\ta_M)$ except that they are independent of the dividend-information Brownian motion $D$.\footnote{We do not need to impose a Gaussian structure on the private information variables because our investors' optimal strategies do not involve filtering. This is because --- as we shall see in \eqref{S0} below --- the equilibria we construct initially reveal a sufficient statistic for how investor rebalancing targets affect price dynamics.} 

We differentiate between two types of targeted investors based on their trading targets $\ta_i$. We refer to traders with targets $\ta_i \neq 0$ as \emph{rebalancers}. These are large asset managers who use dynamic order-execution algorithms to trade on their targets. Traders with $\ta_i =0$ do not need to trade per se but can provide liquidity, and so we call these \emph{intraday liquidity providers}.

The target ratio $\gamma(t)$ is a deterministic function that gives how much of her daily target $\ta_i$ investor $i$ would ideally hold at each time $t$ during the day. The trajectory $\gamma(t) \ta_i$ can be thought of as a time-weighted average price (TWAP) strategy or as, more generally, a variation on an Almgren-Chriss trading strategy.\footnote{It is a ``variation'' in the sense that Almgren and Chriss (1999, 2000) solve for optimal orders given specific risk and cost objectives, whereas we represent investor $i$'s risk and execution-cost objectives  implicitly via the $L_{i,t}$ penalty. } Realistically, the target ratio $\gamma(t)$ should be non-decreasing over the day with $\gamma(1) = 1$ at the end of the day. For example, for a standard TWAP strategy, the target ratio grows linearly with time as $\gamma(t) := t$. Alternatively, for a more generalized strategy, $\gamma(t)$ might follow the shape of the average cumulative volume curve over the trading day. The assumption that $\gamma(t)$ is deterministic  simplifies the analysis. Later, Section \ref{sec:VWAP} extends the model to allow for stochastic target ratios $\gamma_t$ that are then related to VWAP trading in Section \ref{connVWAP}.

\item There are $\overline{M}\in \N$ price-sensitive \emph{realtime hedgers} indexed by $i \in \{M+1,...,M+\overline{M}\}$.  These hedgers have have no private information. Their cumulative penalty $L_{i,1}$ in \eqref{objective} accrues via the process
\begin{align}\label{defLL}
L_{i,t}:=\int_0^t \overline{\kappa}(s)\big(\theta_{i,s}-\epsilon B_s\big)^2ds,\quad t\in[0,1],\quad i \in \{M+1,...,M+\overline{M}\},
\end{align}
which is similar to the common-value specification in Sannikov and Skrzypacz (2016). In \eqref{defLL}, the process  $B = (B_t)_{t\in[0,1]}$ with $B_0=0$ is a standard Brownian motion representing a publicly observable risk-factor that is independent of all other random variables, and $\epsilon\ge0$ scales the hedger target holdings relative to the risk-factor $B_t$.\footnote{For simplicity, we assume $\epsilon$ is constant. Alternatively, it could be a time-dependent function $\epsilon(t)$.} The  hedgers' penalty-severity function $\overline{\kappa}(s)$ is potentially different from $\kappa(s)$ in \eqref{defL}. In the special case $\overline{\kappa}(s):=0$, the hedgers reduce to risk-neutral  \emph{Merton investors} with potential price impact. Another special case is $\epsilon:=0$ in which the hedgers coincide with the intraday liquidity providers discussed above but with a possibly different penalty-severity function $\overline{\kappa}$.

Trading by the hedgers injects intraday randomness (due to $B_t$) in the aggregate trading-demand imbalances over the day.  However, unlike noise traders, the hedgers are price-sensitive and optimize their trades.
\end{itemize}

Investors optimize their trading in actual markets given a variety of considerations. These include both the underlying utility gains from changing their holdings and also bid-ask and other order-execution costs (if taking liquidity) or profits (if providing liquidity), inventory and risk-management costs, and predictable wealth effects due to trading with/against price pressure.  Our analysis decomposes trading optimization for targeted investors into two parts:  First, we interpret the intraday trajectory $\gamma(t) \ta_i$ as a partially optimized strategy given the end-of-day latent target $\ta_i$ and given  order-execution, inventory, and risk-management costs but omitting wealth effects due to price pressure.\footnote{In other words, $\gamma(t) \ta_i$ is the ideal trading trajectory if investor $i$ could trade at prices $D_t$ with just order-execution, risk, and inventory costs but without the price impact of order-flow imbalances.} Second, the targeted investor's optimization problem in \eqref{objective} then adjusts the partially optimized strategy $\gamma(t) a_i$ for price-pressure effects to obtain the fully optimized strategy $\theta_{i,t}$.  In particular, \eqref{objective} includes the expected gain/loss from how the holdings $\theta_{i,t}$ comove with/against price pressure and also an expected penalty, given $L_{i,t}$ in \eqref{defL}, which is a reduced-form representing incremental increased order-execution, inventory, and risk-management costs given the deviation of $\theta_{i,t}$ from the partially optimized trajectory $\gamma(t) \ta_i$ over the day.\footnote{To keep the model parsimonious, we assume rebalancers and liquidity providers have the same penalty-severity function $\kappa(t)$ (where $\gamma(t)$ does not matter for liquidity providers for whom $\ta_i = 0$). In a richer model, the liquidity-provider penalty severity might differ due to additional market-making considerations such as funding costs, risk-aversion, moral hazard costs due to risk limits arising from in-firm principal-agent conflicts, and the fact that liquidity providers are more likely then rebalancers to earn than to pay the bid-ask spread.} An analogous logic justifies the hedger trading optimization \eqref{objective} with the penalty in \eqref{defLL} except that now the partially optimized hedger strategy --- given order-execution, risk-management, and inventory costs and inefficiency costs from incomplete hedging but not wealth effects from price pressure --- is the stochastic process $\epsilon B_t$. 

Given this motivation, there are several specific points to note:  First, the decomposition leading to \eqref{objective} makes our model mathematically tractable, because the quantities $\gamma(t)$, $\epsilon$, $\kappa(t)$, and $\overline{\kappa}(t)$ describing investor trading preferences are inputs in our analysis rather than something to be solved for formally.   

Second, our analysis makes a distinction between price pressure $S_t - D_t$, which is explicitly modeled, and order-execution costs --- like bid-ask spread add-ons, brokerage fees, and exchange fees --- that are implicitly modeled as one of the determinants of the penalties $L_{i,t}$ in \eqref{defL} and \eqref{defLL}.\footnote{The implicit order-execution costs are add-ons on top of $S_t$ for individual transactions of individual investors, whereas $S_t -D_t$ is market-wide price pressure in all transactions for all investors. } In particular, the target ratio function $\gamma(t)$, hedging scalar $\epsilon$, and penalty-severity functions $\kappa(t)$ and $\overline{\kappa}(t)$ all implicitly depend on order-execution costs as well as on inventory and other costs. Intuitively, variation in the target ratio $\gamma(t)$ and penalty severity $\kappa(t)$ over the day reflects, in part, time patterns in order-execution costs (e.g., $U$-shaped average bid-ask spreads, see McInish and Wood 1999, and price impacts, see Barardehi and Bernhardt 2018) and an intensifying inventory-holding preference to reach the target $\ta_i$ towards the end of the day. Our analysis allows, in particular, for penalty-severity functions $\kappa(t)$ that explode towards the end of the trading day as $t \uparrow 1$ as well as for bounded penalty severities. 

Third, the targeted investors and hedgers are all price-sensitive, and so they adjust their stock holdings  $\theta_{i,t}$ relative to their intraday trading targets in response to premia and discounts in prices in order to clear the market over time given the  aggregate latent trading-demand imbalances $\gamma(t)\taS + \bar{M}\epsilon B_t$. In particular, their intraday targets are soft rather than hard constraints. Thus, there are two competing drivers in the objective \eqref{objective}. On one hand, neglecting the penalty term $L_{i,1}$ in  \eqref{objective}, investor $i$ would maximize her expected trading profit from liquidity provision.  On the other hand, neglecting the wealth term $X_{i,1}$ in  \eqref{objective} means investor $i$ would minimize the penalty $L_{i,1}$. In this case, targeted investors  $i\in\{1,...,M\}$ would use the strategy $\gamma(t) \ta_i$, and the hedgers  $i\in\{M+1,...,M+\overline{M}\}$ would hold $\epsilon B_t$. The equilibrium strategies $\hat{\theta}_{i,t}$ in \eqref{optimaltheta_ismall} below strike an optimal balance between these two competing forces where the penalty-severity functions $\kappa(t)$ and $\overline{\kappa}(t)$ determine the relative importance of these two forces over the trading day  $[0,1]$. 

Fourth, an important point about the targeted investors is that they care, not just about their terminal trading targets $\ta_i$, but about the entire path of their holdings over the day relative to their intraday target trajectories $\gamma(t)\ta_i$. In other words, the rebalancers' latent stock demand over the day given $\ta_i \neq 0$ has a time-varying trend controlled by $\gamma(t)$. This aspect of our model --- which allows us to study TWAP and other forms of targeted trading --- is new relative to the previous literature, which has modeled constant zero targets (Du and Zhu 2017), driftless stochastic targets (Vayanos 1999 and Sannikov and Skrzypacz 2016), and fixed terminal targets  but no intraday targets (Brunnermeir and Pedersen 2005). 

Market clearing in the model takes the following form: Recall that all initial stock positions $\theta_{i,-}$ have been normalized to zero. Given an aggregate stock supply normalized to zero, market clearing requires equilibrium investor-holdings $\hat{\theta}_{i,t}$ over the day to satisfy\footnote{It is possible to extend our model to include noise-trader orders such that the floating stock supply becomes an exogenous stochastic process $a(t) + b(t)Z_t +c(t)B_t$ where $a,b$, and $c$ are deterministic functions of time $t\in[0,1]$, $B_t$ is the risk-factor Brownian motion in \eqref{defLL}, and $Z_t$ is a Brownian motion independent of all other random variables.} 
\begin{align}\label{def:clearing_eq}
\sum_{i=1}^{M+\overline{M}}\hat{\theta}_{i,t}=0 \quad \text{for all times }t\in[0,1].%+...+\hat{\theta}_{M,t}+\hat{\theta}_{M+1,t}+\hat{\theta}_{M+\overline{M},t}
\end{align}
The money market is also normalized to be in zero supply.  By Walras law, clearing in the stock market implies clearing in the money market because investors use self-financing strategies. In particular, we assume the initial money-market endowments $\theta^c_{1,-},...,\theta^c_{M+\overline{M},-}$ clear the money market
\begin{align}\label{initialclearing}
%\sum_{i=1}^M \theta_{i,-}=w,\quad
\sum_{i=1}^{M+\overline{M}}\theta^c_{i,-}=0.
\end{align}

 {  The following analysis distinguishes between conceptually different stock-price processes. Let $S = (S_t)_{t\in[0,1]}$ denote a generic stock-price process defined in terms of its  dynamics 
\begin{align}\label{dSS}
\begin{split}
dS_t &:= \mu_t dt + dN_t, \quad S_0:=\hat{S}_0,
\end{split}
\end{align}
where $\mu= (\mu_t)_{t\in[0,1]}$ is a generic drift process, 
 $N= (N_t)_{t\in[0,1]}$ is a fixed martingale, and $\hat{S}_0$ is a fixed initial stock price. } The equilibrium price process  we seek to determine is denoted by $\hat{S}= (\hat{S}_t)_{t\in[0,1]}$ and has two boundary conditions at times $t\in \{0,1\}$:
 
First, the equilibrium stock price $\hat{S}_1$ at the end of the trading day at $t=1$ is pinned down by a reduced-form  end-of-day requirement\footnote{If the terminal restriction \eqref{terminalS} is eliminated, our model becomes simpler because the stock volatility becomes a free parameter and can, for example, be set to be one. The fact that competitive Radner equilibrium models without dividends have free volatilities is well-known; see, e.g., Theorem  4.6.3 in Karatzas and Shreve (1998).} 
\begin{align}\label{terminalS}
\hat{S}_1 =D_1 + \varphi_0 \epsilon B_1+ \varphi_1 \taS 
\end{align}
where $\varphi_0,\varphi_1\in \R$ are exogenous constants and the total target imbalance is denoted by%\footnote{The target imbalance $\taS$ is non-zero due to rebalancing since liquidity providers have targets $\ta_i=0$.}
\begin{align}
 \taS:= \sum_{i=1}^M\tilde{a}_i\label{taS1}.
\end{align}
%{ Conditional on $\taS$, the terminal equilibrium stock price \eqref{terminalS} is Gaussian, which is a common feature in many price impact equilibrium models (see, e.g., Kyle 1985 and Grossman and Stiglitz 1980).} 
When our model is applied to a short time horizon (e.g., a trading day), the end-of-day price $\hat{S}_1$ is assumed to be the overnight valuation that clears the stock market given the latent trading-demand imbalances due to the aggregate rebalancer and hedger targets. { In particular, market-clearing at $t = 1$ at the end of the day is assumed to reflect overnight stock-holding by the rebalancers, market-makers, and hedgers and also possibly additional net stock demand from overnight liquidity providers who do not trade during the day. 
}If $\varphi_1 > 0$, then a positive aggregate latent target $\ta_\Sigma$ pushes up the end-of-day price $\hat{S}_1$ relative to $D_1$ in order for markets to clear. Similarly, if $\varphi_0>0$, then a positive hedging target $\epsilon B_1$ also raises $\hat{S}_1$ relative to $D_1$. { Both effects are qualitatively natural for how latent trading-demand imbalances might affect end-of-day prices.
} A special case of \eqref{terminalS} is
\begin{align}\label{S_1}
\hat{S}_1 = D_1.
\end{align}
This case applies if $D_1$ is a liquidating dividend paid at time $t=1$ (as in, e.g., Grossman and Stiglitz 1980 and Kyle 1985) or, alternatively, if there are no overnight liquidity effects  {in pricing } at $t=1$. 
 For $(\hat{S}_t)_{t\in[0,1]}$ to satisfy \eqref{terminalS} or \eqref{S_1}, the price dynamics in \eqref{dSS} must be restricted as time approaches maturity (i.e., as $t\uparrow 1$). { As we shall see in Theorem 
\ref{thm:Main} below, the equilibrium stock-price process is linear in $(D_t,\tilde{a}_\Sigma,B_t)$ with time-varying deterministic coefficients. The terminal restriction \eqref{terminalS} gives boundary conditions for these time-varying coefficient functions. 
This allows us to derive endogenous intraday price and investor holding processes that are consistent with the assumed end-of-day price in \eqref{terminalS}.

Second, all investors act as price-takers at time $t=0$. This means the initial price $S_0=\hat{S}_0$ in \eqref{dSS} is unaffected by individual investor holdings $\theta_{i,0}$. This is for tractability. In addition, in equilibrium, the initial orders of investors cause the endogenous opening price $\hat S_0$ to adjust to clear the market given the underlying latent trading-demand imbalance due to the aggregate target $\ta_\Sigma$. From a modeling perspective, the initial stock price $\hat{S}_0$ at time $t=0$ is required to reveal the aggregate target $\taS$ defined in \eqref{taS1} in the sense that 
\begin{align}\label{S0}
\sigma(\hat{S}_0) =  \sigma(\taS).
\end{align}
The measurability requirement \eqref{S0} allows investors to avoid filtering and thereby keeps the model tractable. }

The information structure of our model is as follows: For tractability, all traders have homogeneous beliefs in the sense that they all believe the processes $(D,B)$ are the same independent Brownian motions. {At each time  $t\in[0,1]$ over the trading day}, the dividend-information process $D_t$  and the real-time hedging factor  $B_t$ are publicly observed. At each time $t\in[0,1]$, each investor $i$ chooses a stock-holding position $\theta_{i,t}$ adapted to the filtration\footnote{As usual in continuous-time models, certain stochastic integrals need to be martingales; see Definition \ref{def_admis} below for details. Also, as usual, we have implicitly augmented \eqref{meas1} with null-sets to ensure that the ``usual conditions" hold (see, e.g., Protter 2004 for details).}
\begin{align}\label{meas1}
\begin{split}
\sF_{i,t}:=
\begin{cases}
\sigma(S_u,D_u,B_u,\tilde{a}_i,\theta^c_{i,-})_{u\in[0,t]}, \quad i=1,...,M,\\
\sigma(S_u,D_u,B_u,\theta^c_{i,-})_{u\in[0,t]}, \quad i=M+1,...,M+\overline{M}.
\end{cases}
\end{split}
\end{align}  
While the risk factor $B_t$ is observable, { the requirement $S_0 = \hat{S}_0$ in \eqref{dSS} and the measurability property \eqref{S0} ensure the total latent target $\ta_\Sigma$ can be inferred from $S_0$.}

\section{Individual optimization problems}\label{sec:22}

This section gives a precise description of the optimization problem in  \eqref{objective} for the targeted investors and hedgers.  Our analysis proceeds in two steps: First, we describe technical properties of the set of admissible holding strategies $\sA_i$ in  \eqref{objective}.
Second, we construct perceptions of off-equilibrium market-clearing prices for a generic investor $i$ in \eqref{objective} when using an arbitrary holding strategy $\theta_i\in \sA_i$. Section \ref{sec:2} then derives equilibrium strategies $\hat{\theta}_{i,t}$ for investor $i$ given these price beliefs. \ \\

\noindent {\bf Step 1:} The set of admissible strategies $\sA_i$ for investor $i$ in \eqref{objective} is defined as follows where a c\`agl\`ad process is a left-continuous process with right limits:

\begin{definition}\label{def_admis}  For a given stock-price process $S_t$ with drift $\mu_t$ and martingale $N_t$ as in \eqref{dSS} with a predictable quadratic variation process $\langle N\rangle_t$, we define $\sF_{i,t}$ by \eqref{meas1}. A c\`agl\`ad process $\theta_{i}= (\theta_{i,t})_{t\in[0,1]}$ adapted to $\sF_{i,t}$ is \emph{admissible}, and we write $\theta_{i}\in\sA_i$, if the following integrability condition holds:
\begin{align}\label{admis_integrability}
\E\left[\int_0^1 \big(|\theta_{i,t}\mu_t|dt+ \theta_{i,t}^2 d\langle N\rangle_t\big)\,\Big|\,\sF_{i,0}\right]<\infty,\quad i\in\{1,...,M+\overline{M}\}.
\end{align}
$\endproof$
\end{definition}
\noindent It is well-known that an integrability condition like \eqref{admis_integrability} rules out doubling strategies (see, e.g., Chapters 5 and 6 in Duffie 2001). The equilibrium stock-holding process in \eqref{optimaltheta_ismall} below is not bounded (because it depends linearly on $B_t$) but does still satisfy \eqref{admis_integrability}. There are two reasons for the left-continuity requirement placed on $\theta_{i,t}$. First, left-continuity of the investor holding paths is sufficient { to ensure that the state-process $Y_t$ appearing in \eqref{dXj} below is left-continuous (ultimately, $Y_t$ in \eqref{dXj} becomes the solution $Y_t^{\theta_i}$ to the stock-market clearing equation \eqref{def:clearing_eq0}). In turn, $Y_t$'s left-continuity is sufficient for investors to infer $Y_t$ from past and current stock-price observations (see Lemma \ref{lem:respone} in Appendix \ref{app:proofs})}. Second, in Example \ref{ex:gamma} in Section \ref{sec:VWAP}, which is an extension of our model to stochastic targets, the martingale $N_t$ is only c\`adl\`ag (i.e., right-continuous with left limits). When $N_t$ has such points of discontinuity, left-continuity of $\theta_{i,t}$ and the integrability condition \eqref{admis_integrability} are sufficient to ensure martingality of the stochastic integral $\int_0^t \theta_{i,u}dN_u$, $t\in[0,1]$ (see, e.g., p.171 in Protter 2004).\footnote{Verification of optimality of conjectured optimizers in continuous-time stochastic control problems always involves proving certain stochastic integrals are martingales. This is illustrated in Duffie (2001, Chapter 9C).} Except for Example \ref{ex:gamma}, however, strategies $\theta_{i,t}$ are continuous and, thus, are automatically left-continuous in Definition \ref{def_admis}.  \ \\

\noindent {\bf Step 2:} This step constructs perceptions for a generic investor $i$ about the market-clearing prices $S_t^{\theta_i}$ she faces given arbitrary holdings $\theta_{i,t}$.  In equilibrium, price dynamics and price perceptions must agree. Off equilibrium, however, price perceptions must be reasonable. We describe two different cases of reasonable off-equilibrium price beliefs.  One case is standard pricing-taking.  This is the simplest version of our model. The second case has price-impact in the sense that investors are strategic and believe their individual holdings affect off-equilibrium market-clearing prices. { In particular, although we assume investors are price-takers at the market open at $t = 0$, our second case allows for price impact of investor holdings $\theta_{i,t}$ over the rest of the day for $t \in (0,1)$. }\ \\

\noindent {\bf Case of price-taking perceptions:} Optimal holdings for investors with price-taking beliefs are derived from \eqref{objective} under the assumption that the off-equilibrium prices $S^{\theta_i}_t$ { each investor $i$ perceives herself} as facing, the martingale $N_t$, and the perceived price drift $\mu^{\theta_i}_t$ given  holdings $\theta_{i,t}$ for investor $i$ are all unaffected by her arbitrary holdings $\theta_{i,t}$.  The equilibrium construction for this case in Sections \ref{sec:2} and \ref{sec:Radner} then derives an equilibrium price process such that the market clears given investor optimal holdings given the equilibrium price process with  price-taking beliefs.\ \\

\noindent {\bf Case of price-impact perceptions:} There are many ways to model price-impact. For example, stock prices are affine functions of investor orders in the discrete-time models in Vayanos (1999, Eq. (4.2)) and Kyle (1985, Eq. (3.12)). A discrete-time version of our model also takes price changes $\Delta S_{t_n}$ as affine functions of the holdings $\theta_{i,t_n}$. However, modeling off-equilibrium prices in continuous-time as functions of the levels of arbitrary holdings over time is technically difficult because, while $\theta_{i,t}$ can serve as an integrand, integration with respect to $d\theta_{i,t}$ might be ill-posed.  Thus, in our continuous-time model of price impact, we { assume} that the perceived price drift $\mu_t^{\theta_i}$ is an affine function of the holdings $\theta_{i,t}$, but that the martingale $N_t$ in \eqref{dSS} and the initial value $S_ 0 = \hat S_0$ are both independent of the holding process $\theta_{i,t}$.  Thus, our model of price-impact is an affine drift specification in the general continuous-time price-impact setting in  Cvitani\' c and Cuoco (1998). 
{ Given the perceived off-equilibrium price impacts, we derive optimal holding strategies for investors in our model.}

Our eventual goal is to construct a Subgame Perfect Nash equilibrium, so the off-equilibrium prices associated with off-equilibrium holdings by a given investor $i$ must be consistent with off-equilibrium beliefs for other investors such that, given the other investors' optimal off-equilibrium holdings given their beliefs, the market clears to produce the prices perceived by investor $i$. We denote the other investors who respond to off-equilibrium holdings by investor $i$ using an index $j \neq i$.  Next, we construct perceived market-clearing prices for investor $i$ and the associated perceived responses $\theta_{j,t}$ of other investors $j\neq  i$ to arbitrary holdings $\theta_{i,t}$ by investor $i$.

Our derivation of optimal investor holding strategies involves two different off-equilibrium stock-price processes in the filtrations in \eqref{meas1} and in the investor wealth dynamics in \eqref{dX}.  First, let $S_t^Y$ denote the off-equilibrium prices as perceived by a generic  investor $j$ with $j \neq i$ given investor $j$'s own holdings $\theta_{j,t}$ and given a general state-process $Y_t$ representing the {perceived} net effect of arbitrary holdings by other investors.\footnote{The filtration $\sF_{i,t}$ in \eqref{meas1} also depends on $S$ and thus $\sF_{i,t}$ varies depending on which stock-price process $S\in \{\hat{S}, S^{\theta_i}, S^Y\}$ we consider in the various steps of the equilibrium construction.}  Second, let $S_t^{\theta_i}$  denote off-equilibrium prices perceived by investor $i$ given investor $i$'s holdings $\theta_{i,t}$ and given the state-process $Y_t^{\theta_i}$ induced by the market-clearing responses of investors $j\neq i$ to $i$'s holdings $\theta_{i,t}$. The difference between these two perceived price processes is that $S_t^Y$ is for an arbitrary exogenous state-process $Y_t$ whereas $S_t^{\theta_i}$ is for an endogenous state-process $Y_t^{\theta_i}$ given the effect of $\theta_{i,t}$ on market-clearing. The prices $S_t^Y$  and $S_t^{\theta_i}$ are linked with each other and with the equilibrium prices $\hat{S}_t$. First, the initial stock prices $S^Y_0$ and $S^{\theta_i}_0$ are always pinned down to be $\hat{S}_0$ in \eqref{dSS}. Thus, all investors act like price takers at time $t=0$. However, the perceived off-equilibrium terminal prices $S^{\theta_i}_1$ and $S^Y_1$ are not required to satisfy the equilibrium terminal condition \eqref{terminalS}. This is permissible because these are perceived off-equilibrium prices that can differ from equilibrium prices. Second, the equilibrium holdings $\hat{\theta}_{i,t}$ for all traders $i\in\{1,..,M+\overline{M}\}$ must produce the same equilibrium stock price $S_t^{\hat{\theta}_i}=\hat{S}_t$, which in turn must satisfy condition \eqref{terminalS} at time $t=1$.

The aspect of price beliefs that matters for modeling the impact of off-equilibrium holdings on market-clearing is how $\theta_{i,t}$ for an investor $i$ at time $t$ affects the market-clearing stock-price drift $\mu_t$ and, thus, the investment attractiveness of holding stock at time $t$. Thus, we start this step with an { assumption } that the perceived general price process $S^Y = (S^Y_t)_{t\in[0,1]}$ for other investors $j \neq i$ in trader $j$'s optimization problem \eqref{objective} has the form in \eqref{dSS} where the perceived stock-price drift $\mu = \mu^Y$ is defined as\footnote{The equilibrium construction in  Section \ref{sec:2} explicitly solves for perceived prices with this conjectured form that are consistent with equilibrium. } 
\begin{align}\label{dXj}
\begin{split}
\mu^Y_t &:=  
\begin{cases}
\nu_0(t) Y_{t} +\nu_1(t)\gamma(t)\taS+\nu_2(t)B_t+\nu_3(t)\theta_{j,t}+ \nu_4(t)\gamma(t)\ta_j ,\quad j\in\{1,...,M\},\\
 \bar{\nu}_{0}(t)Y_{t} +\bar{\nu}_1(t)\gamma(t)\taS+ \bar{\nu}_2(t)B_t+\bar{\nu}_3(t)\theta_{j,t},\quad j\in\{1+M,...,M+\overline{M}\},
\end{cases}
\end{split}
\end{align}
where $Y=(Y_t)_{t\in[0,1]}$ is an exogenous state-process with c\`agl\`ad paths and $\nu_0(t),...,\nu_4(t)$ and $\bar{\nu}_0(t),...,\bar{\nu}_3(t)$ are continuous deterministic functions with $\nu_0(t)\neq0$, $\bar{\nu}_0(t)\neq0$ such that
\begin{align}\label{SOC}
\nu_3(t) < \kappa(t)\;\;\text{when $M\ge1$ and} \;\; \bar{\nu}_3(t) < \overline{\kappa}(t) \;\;\text{when $\overline{M}\ge1$},\quad t\in[0,1).
\end{align} 
The martingale $N$ in the price process \eqref{dSS} for $S^Y$ has dynamics 
\begin{align}\label{dNt}
dN_t:=dD_t + \zeta(t)\epsilon dB_t,\quad N_0:=0,
\end{align}
where $\zeta(t)$ is a continuous deterministic function of time for $t\in[0,1]$.

The coefficients $\nu_0(t), ...,\bar{\nu}_3(t)$ and $\zeta(t)$ in \eqref{dXj} and \eqref{dNt} describe the subjective perceptions (i.e., off-equilibrium beliefs) of investor $j$ about the price process she faces.  We assume all investors perceive the same pricing coefficients. The coefficients $\nu_3(t)$ and $\bar{\nu}_3(t)$ represent, specifically, {an investor's} perception of how her own holdings $\theta_{j,t}$ directly affect prices off-equilibrium. In the discussion here, the perceived price coefficients are taken as given. However, they are endogenized in the equilibrium construction in Section \ref{sec:2}.

Because the state-process $Y_t$ is general (e.g., $Y_t$ need not be Markovian), it is crucial for tractability that investors have linear utilities as in \eqref{objective}. With linear preferences, investors' optimal holding decisions at each time $t\in[0,1]$ only depend on the perceived price drift and the associated holding penalties accruing at that time $t$. Thus, investors optimize pointwise given the impact their holdings have on the perceived price drift.\footnote{More specifically, investors maximize the expectation of the integrand of a Riemann integral of a quadratic function of $\theta_{i,t}$ at each time $t\in[0,1]$. See Eqs. \eqref{resp1} and \eqref{resp2} in  Appendix \ref{app:proofs}. }\ \footnote{An optimal trading strategy is typically computed as the solution to a Hamilton-Bellman-Jacobi equation that takes intertemporal trade-offs into account.  However, in our model the intraday target trajectory $\gamma(t) \ta_i$ and hedge positions $\epsilon B_t$ are interpreted as the solutions to the targeted investors' and hedgers' intertemporal partial optimization problems with order-execution, risk, and inventory costs but when price pressure from aggregate order-flow imbalances is ignored.  As a result, the full optimization problem in \eqref{objective} involves a series of separable trade-offs between incremental penalties for deviations of $\theta_{i,t}$ from the target trajectory and hedge positions at each time $t \in [0,1]$ and the contemporaneous price drift (expected capital gain) at time $t$  when investors adjust their position to  trade on price pressure.}  These arguments lead to Lemma \ref{lem:respone} in Appendix \ref{app:proofs}, which gives the optimal response holdings $\theta^Y_{j,t}$ in \eqref{NASHH1} for targeted investors and hedgers given a perceived price process for investor $j$ of the form described in \eqref{dXj} { through} \eqref{dNt}.\footnote{In Lemma \ref{lem:respone} in Appendix \ref{app:proofs} with price-impact, the state-process $Y_t$ must be inferable, which is ensured by $Y_t$ having left-continuous paths, which follows from investor holdings having left-continuous paths as required by Definition \ref{def_admis}. If investor strategies are continuous, then the state-process $Y_t$ can be taken to be continuous too.}   The restrictions in \eqref{SOC} ensure the second-order condition for optimality of \eqref{NASHH1} is satisfied.

The optimal-response holdings $\theta_{j,t}^Y$ in  Lemma \ref{lem:respone} in Appendix \ref{app:proofs} are for an arbitrary left-continuous state-process $Y_t$. However, to construct perceived prices for investor $i$ that also clear the stock market, the price perceptions of investor $i$ must take into account the fact that her holdings $\theta_{i,t}$ also affect the state-process $Y_t$ perceived by other investors $j\neq i$ via the market-clearing condition:
\begin{align}\label{def:clearing_eq0}
\theta_{i,t} + \sum_{j\neq i,j=1}^{M+\overline{M}} \theta^Y_{j,t}=0,\quad t\in[0,1],
\end{align}
where $\theta^Y_{j,t}$ denotes optimal responses of investors $j\neq i$ in \eqref{NASHH1} in Lemma \ref{lem:respone} in Appendix \ref{app:proofs} to a state-process $Y_t$. Thus, given arbitrary holdings $\theta_{i,t}$, we can then solve \eqref{def:clearing_eq0} for the associated state-process, which we denote by $Y^{\theta_i}_t$. The state-process $Y^{\theta_i}_t$ ensures that the off-equilibrium optimal-response holdings for traders $j\in\{1,...,M+\overline{M}\}\setminus\{i\}$ clear the stock market as trader $i$ varies her off-equilibrium holdings $\theta_{i,t}$. The perceived off-equilibrium market-clearing stock-price process $S^{\theta_i}_t$ associated with  $\theta_{i,t}$ has a drift process $\mu^{\theta_i}_t$ given by \eqref{dXj} with $Y_t:=Y^{\theta_i}_t$. However, the initial price $S^{\theta_i}_0:=\hat{S}_0$ and martingale $N_t$ in $dS^{\theta_i}_t$ in \eqref{dSS} do not depend on $\theta_{i,t}$.

%Using the perceived price process above with the endogenous state process $Y^{\theta_i}_t$ determined by the market clearing condition (2.6) leads to the following characterization of the perceived price process for a representative investor $i$:

\begin{lemma}\label{lem:new} {Assume that  $\nu_0(t)\neq0$, $\bar{\nu}_0(t)\neq0$, $t\in[0,1]$, and that \eqref{SOC} holds}. The off-equilibrium stock-price process perceived by investor $i\in \{1, ..., M + \overline{M}\}$ for arbitrary holdings $\theta_{i}\in \sA_i$ {is defined by}
\begin{align}\label{Smu}
S^{\theta_i}_t &:= \hat{S}_0 + \int_0^t \mu_u^{\theta_i}du +N_t,\quad t\in[0,1],
\end{align}
where the stock-price drift is defined by
\begin{footnotesize}
\begin{align}\label{dSmu8}
\begin{split}
\mu^{\theta_i}_t &:= 
\begin{cases}
\frac{2 \kappa  \nu_0 (\bar{\nu}_3-\overline{\kappa}) +(\bar{\nu}_3-\overline{\kappa} )\nu_0 \nu_4 +\overline{M} (\kappa -\nu_3) (\nu_1 \bar{\nu}_0-\nu_0 \bar{\nu}_1)}{(M-1) \nu_0 (\overline{\kappa}-\bar{\nu}_3)+\overline{M} \bar{\nu}_0 (\kappa -\nu_3)} \gamma\taS-\frac{\overline{M} (\kappa -\nu_3) (\nu_0 \bar{\nu}_2-\nu_2 \bar{\nu}_0+2 \overline{\kappa} \nu_0 \epsilon )}{(M-1) \nu_0 (\overline{\kappa}-\bar{\nu}_3)+\overline{M} \bar{\nu}_0 (\kappa -\nu_3)}B_t
\\
\quad +\frac{\kappa  (-2 \overline{\kappa} \nu_0+2 \nu_0 \bar{\nu}_3+\nu_3 \overline{M} \bar{\nu}_0)-\nu_3 ((M+1) \nu_0 (\bar{\nu}_3-\overline{\kappa})+\nu_3 \overline{M} \bar{\nu}_0)}{(M-1) \nu_0 (\overline{\kappa}-\bar{\nu}_3)+\overline{M} \bar{\nu}_0 (\kappa -\nu_3)}\theta_{i,t}
\\
\quad +\frac{2  \kappa  \nu_0 (\overline{\kappa}-\bar{\nu}_3)+M \nu_0 \nu_4 (\overline{\kappa}-\bar{\nu}_3)+\nu_4 \overline{M} \bar{\nu}_0 (\kappa -\nu_3)}{(M-1) \nu_0 (\overline{\kappa}-\bar{\nu}_3)+\overline{M} \bar{\nu}_0 (\kappa -\nu_3)} \gamma\ta_i, \quad i\in\{1,...,M\},
\vspace{0.1cm}
\\
-\frac{(\overline{\kappa}-\bar{\nu}_3) (\bar{\nu}_0 (2  \kappa +M \nu_1+\nu_4)-M \nu_0 \bar{\nu}_1)}{M \nu_0 (\overline{\kappa}-\bar{\nu}_3)+(\overline{M}-1) \bar{\nu}_0 (\kappa -\nu_3)}\gamma\taS
-\frac{M (\overline{\kappa}-\bar{\nu}_3) (\nu_2 \bar{\nu}_0-\nu_0 \bar{\nu}_2)+2 \overline{\kappa} (\overline{M}-1) \bar{\nu}_0 \epsilon  (\kappa -\nu_3)}{M \nu_0 (\overline{\kappa}-\bar{\nu}_3)+(\overline{M}-1) \bar{\nu}_0 (\kappa -\nu_3)}B_t
\\
\quad +\frac{2 \overline{\kappa} \bar{\nu}_0 (\nu_3-\kappa )+\overline{\kappa} M \nu_0 \bar{\nu}_3-M \nu_0 \bar{\nu}_3^2+(\overline{M}+1) \bar{\nu}_0 \bar{\nu}_3 (\kappa -\nu_3)}{M \nu_0 (\overline{\kappa}-\bar{\nu}_3)+(\overline{M}-1) \bar{\nu}_0 (\kappa -\nu_3)}\theta_{i,t}, \quad i\in\{M+1,...,M+\overline{M}\},
\end{cases}
\end{split}
\end{align} 
\end{footnotesize}and where the martingale $N_t$ is as in \eqref{dNt}, and given an initial stock price  $\hat{S}_0$ that satisfies \eqref{S0}. The price process \eqref{Smu} clears the stock market in the sense that \eqref{def:clearing_eq0} holds.
\end{lemma}

We note three consequences of Lemma \ref{lem:new}: First, because the state-process $Y^{\theta_i}_t$ in \eqref{def:clearing_eq01} in Appendix \ref{app:proofs} is affine in $\theta_{i}\in \sA_i$, we see that $Y^{\theta_i}_t$ has left-continuous paths because $\theta_{i,t}$ has left-continuous paths (see Definition \ref{def_admis}). Furthermore, from \eqref{NASHH1} in Lemma \ref{lem:respone} in Appendix \ref{app:proofs}, investor $j$'s optimal response $\theta^{Y^{\theta_i}}_{j,t}$  for $j\neq i$ is also affine in trader $i$'s  off-equilibrium holdings $\theta_{i,t}$.  Second, because the price drift \eqref{dSmu8} is affine in $\theta_{i,t}$, the corresponding optimization problem \eqref{objective} is a quadratic problem when the stock price is defined as in \eqref{Smu}. These two properties are used to derive the equilibrium holdings $\hat{\theta}_{i}\in \sA_i$ in Theorem \ref{thm:Main} in Section \ref{sec:2}. Third, the case of price-taking perceptions is a special case of the price-impact case in which the coefficients multiplying $\theta_{i,t}$ in \eqref{dSmu8} are zero. Our equilibrium construction in Section \ref{sec:2} is for the general price-impact case.  Section \ref{sec:Radner} then derives conditions on the price-perception coefficients in \eqref{dXj} for equilibrium with price-taking perceptions.

\section{Equilibria with deterministic targets}\label{sec:2}
Having described the individual investor optimization problems in Section \ref{sec:22}, this section formally defines and constructs an equilibrium.

\begin{definition}[Equilibrium]\label{def_eq} A \emph{Subgame Perfect  Nash equilibrium} consists of perceived price coefficients given by deterministic functions $\nu_0(t),...,\nu_4(t)$, and $\bar{\nu}_0(t),...,\bar{\nu}_3(t)$ in \eqref{dXj} with $\nu_0(t)\neq0$ and $\bar{\nu}_0(t)\neq0$, an initial stock price $\hat{S}_0$ { satisfying \eqref{S0}}, and a martingale $N= (N_t)_{t\in[0,1]}$ such that, given the perceived stock-price process \eqref{Smu}, the resulting optimal stock-holding processes $\hat{\theta}_{1,t},...,\hat{\theta}_{M+\overline{M},t}$ from \eqref{objective} satisfy the following conditions:
\begin{itemize}

\item[(i)] The optimal holdings $\hat{\theta}_{i}\in \sA_i$, $i\in\{1,...,M+\overline{M}\}$, satisfy the intraday market-clearing condition \eqref{def:clearing_eq}.
\item[(ii)] When $\theta_{i,t}$ is set to the optimizer $\hat{\theta}_{i,t}$ in \eqref{dSmu8}, the resulting stock-price drifts $\mu^{\hat{\theta}_i}_t$ are the same $\hat{\mu}_t$ for all investors $i \in\{1,...,M+\overline{M}\}$. The corresponding equilibrium stock-price process from \eqref{Smu} with $\theta_{i,t}=\hat{\theta}_{i,t}$ is denoted by $\hat{S}_t$.
\item[(iii)]  The stock-price process $\hat{S}_t$ satisfies the terminal price condition \eqref{terminalS} at time $t=1$
for given constants $\varphi_0,\varphi_1\in\R$.
\end{itemize}
$\endproof$
\end{definition}
\noindent { Our equilibrium concept is stronger than Nash because it involves beliefs for each investor $i$ about the perceptions of other investors $j \ne i$ that determine investor $j$'s optimal responses to hypothetical off-equilibrium holdings by investor $i$.
}

Our main result is Theorem \ref{thm:Main} below, which gives restrictions ensuring  equilibrium existence (proof is in Appendix \ref{app:proofs}). As we shall see, there are two degrees of freedom in the perceived price coefficients $\nu_0(t),...,\nu_4(t)$ and $\bar{\nu}_0(t),...,\bar{\nu}_3(t)$, and so there are multiple (indeed, infinitely many) equilibria. This situation is similar to Vayanos (1999, Sec. 5) and Sannikov and Skrzypacz (2016).  Keeping the price impact coefficients $\nu_3(t)$ and $\bar{\nu}_3(t)$ as the two free parameters simplifies the exposition. In Section \ref{sec:Radner} we use the mathematical flexibility of these free functions to consider equilibria with different amounts of competition and strategic behavior. 

The equilibrium stock-price process $\hat{S}_t$ will be shown to have the form
\begin{align}\label{S_equilibrium}
\hat{S}_t:= D_t +g(t)\tilde{a}_{\Sigma}+ \zeta(t)\epsilon B_t,
\end{align}
for two continuously differentiable deterministic functions $g,\zeta:[0,1]\to\R$.
Consequently, the equilibrium stock-price drift $\hat{\mu}_t$ and martingale $N_t$ in \eqref{dSS} are given by 
\begin{align}\label{S_equilibriumm}
\begin{split}
\hat{\mu}_t &=  g'(t)\tilde{a}_{\Sigma}+   \zeta'(t)\epsilon B_t,\\ 
dN_t&=dD_t + \zeta(t)\epsilon dB_t,\quad N_0=0.
\end{split}
\end{align}
We note three features about \eqref{S_equilibrium}. First, equilibrium prices are expressed in \eqref{S_equilibrium} as functions of the underlying trading-demand variables $\ta_\Sigma$ and $B_t$.  In particular, prices are functions of market-clearing investor holdings, which are functions of the underlying latent trading-demand variables $\ta_{i}$ (which aggregate to $\ta_\Sigma$) and $B_t$. The intuition is that equilibrium prices depend on the underlying latent total trading demand, which includes both trades that occur in equilibrium and also trading-demand imbalances that prices deter so that markets clear. Making the role of latent trading demand --- and especially demand imbalances due to intraday TWAP trading targets --- explicit is one of the contributions of our analysis. Second, given $\tilde{a}_{\Sigma}$, the equilibrium stock-price process is Gaussian. More specifically, the price process \eqref{S_equilibrium} is a Bachelier model with time-dependent coefficients.\footnote{While the equilibrium stock price can be negative with positive probability, such Gaussian models have been widely used in the  market microstructure literature by, e.g., Grossman and Stiglitz (1980) and Kyle (1985). Gaussian models are also widely used in the optimal order-execution literature including   Almgren and Chriss (1999, 2001); see, e.g., the discussion in the Gatheral and Schied (2013, Section 3.1) survey.} Third, the individual investors' perceived price processes in \eqref{Smu} exhibit path dependency in the sense that the path of $(B_u)_{u\in[0,t]}$ is needed to determine $S^{\theta_i}_t$ for $t\in[0,1]$ for an arbitrary holding process $\theta_{i,t}$. However, only the current value { of $B_t$ affects} the equilibrium prices $\hat{S}_t$ in \eqref{S_equilibrium}.

\begin{theorem}\label{thm:Main}  Let $\gamma:[0,1]\to [0,\infty)$ be a continuous function, let $\nu_3,\bar{\nu}_3:[0,1) \to \R$ be continuous functions, let $\kappa,\overline{\kappa} :[0,1) \to (0,\infty)$ be continuous and integrable functions; i.e.,
\begin{align}\label{square_integrable}
\int_0^1 \big(\kappa(t)+\overline{\kappa}(t)\big)dt <\infty,
\end{align}
that satisfy the second-order conditions \eqref{SOC}, and let there be at least $M+\overline{M}\ge2$ investors. In addition, assume that \eqref{initialclearing} holds in the money market. Define the functions in \eqref{S_equilibrium} by
\begin{align}\label{ODEs_equilibrium}
\begin{split}
g(t)&:=\varphi_1-\int_t^1 \mu_1(u)\gamma(u)du,\\
\zeta(t) &:=\varphi_0  -\int_t^1 \mu_2(u)du,
\end{split}
\end{align}
for $t\in[0,1]$ where\footnote{For notational brevity, the time arguments for $\mu_1(t),\mu_2(t),\kappa(t),\overline{\kappa}(t),\nu_3(t)$, and $\bar{\nu}_3(t)$ are suppressed.}
\begin{footnotesize}
\begin{align} \label{mu12} 
\begin{split}
\mu_1&:= 
-\frac{2\kappa\big(2(M+\overline{M})\overline{\kappa}-(1+M+\overline{M})\bar{\nu}_3\big)}{M\big(2(M+\overline{M})\overline{\kappa}-(1+M+\overline{M})\bar{\nu}_3\big) +\overline{M} \big(2(M+\overline{M})\kappa - (1+M+\overline{M})\nu_3\big)},\\
\mu_2&:= -\frac{2\overline{M}\overline{\kappa}\big(2(M+\overline{M})\kappa - (1+M+\overline{M})\nu_3\big)}{M\big(2(M+\overline{M})\overline{\kappa}-(1+M+\overline{M})\bar{\nu}_3\big) +\overline{M} \big(2(M+\overline{M})\kappa - (1+M+\overline{M})\nu_3\big)}.
\end{split}
\end{align}
\end{footnotesize}Provided that $g(0)\neq0$, price-perception functions $\nu_0\neq 0,\nu_1,\nu_2,\nu_4$, and $\bar{\nu}_0,\bar{\nu}_1,\bar{\nu}_2$ satisfying \eqref{nubar0} and \eqref{nu4} in Appendix \ref{app:proofs} together with $N_t$ in \eqref{S_equilibriumm} form an equilibrium in which:
\begin{itemize}
\item[(i)] Investor equilibrium holdings are given by 
\begin{align} \label{optimaltheta_ismall} 
\hat{\theta}_{i,t} =
\begin{cases}
\alpha_1(t)\gamma(t)\taS+\alpha_2(t)\epsilon B_t+\alpha_3(t)\gamma(t) \ta_i,\quad i\in\{1,...,M\},\\
 \bar{\alpha}_1(t)\gamma(t)\taS+ \bar{\alpha}_2(t)\epsilon B_t,\quad i\in\{M+1,...,M+\overline{M}\},
\end{cases}
\end{align}
where
\begin{align}\label{optimaltheta_alphas}
\begin{split}
\alpha_1&:= \frac{(M+\overline{M}-1)\mu_1}{2(M+\overline{M})\kappa - (1+M+\overline{M})\nu_3},\\
\alpha_2&:=  \frac{(M+\overline{M}-1)\mu_2}{2(M+\overline{M})\kappa - (1+M+\overline{M})\nu_3},\\
\alpha_3&:= \frac{2\kappa(M+\overline{M}-1)}{2(M+\overline{M})\kappa - (1+M+\overline{M})\nu_3},\\
\bar{\alpha}_1 &=  \frac{(M+\overline{M}-1)\mu_1}{2(M+\overline{M})\overline{\kappa} - (1+M+\overline{M})\bar{\nu}_3},\\
\bar{\alpha}_2&:= - \frac{M(M+\overline{M}-1)\mu_2}{\overline{M}\big(2(M+\overline{M})\kappa - (1+M+\overline{M})\nu_3\big)}.
\end{split}
\end{align}

\item[(ii)]  The equilibrium stock price $\hat{S}_t$ is { given by \eqref{S_equilibrium} with $g(t)$ and $\zeta(t)$ from \eqref{ODEs_equilibrium}. The associated equilibrium price drift in \eqref{S_equilibriumm} is }
\begin{align} \label{dSmuuuu} 
\hat{\mu}_t &:= \mu_1(t)\gamma(t)\taS+ \mu_2(t)\epsilon B_t,
\end{align}
{ with  $\mu_1(t)$ and $\mu_2(t)$ from \eqref{mu12}.}
\end{itemize}
\end{theorem}

\begin{remark}\label{rmkthm} We note several properties of this equilibrium here:

\begin{enumerate}

\item From \eqref{S_equilibrium}, the initial {equilibrium} stock price at $t=0$ is
\begin{align}\label{S_000}
\begin{split}
\hat{S}_0 &= D_0+ g(0) \taS+ \zeta(0)\epsilon B_0\\
&= D_0+g(0) \taS,
\end{split}
\end{align}
where the second equality follows because $B_0=0$. Therefore, whenever $g(t)$ from \eqref{ODEs_equilibrium} satisfies 
$g(0)\neq0$, the aggregate target $\taS$ can be inferred from $\hat{S}_0$ and vice  versa given that $D_0$ is public information. Thus, when $g(0)\neq0$, we have  $\sigma(\hat{S}_0)= \sigma(\taS)$ as required in \eqref{S0}. By \eqref{ODEs_equilibrium}, the condition $g(0)\neq0$ is equivalent to
\begin{align}
\varphi_1\neq\int_0^1 \mu_1(u)\gamma(u)du,
\end{align}
where $\mu_1$ is from \eqref{mu12}. Consequently, for given functions $(\gamma,\kappa,\overline{\kappa},\nu_3,\bar{\nu}_3)$ satisfying the second-order conditions \eqref{SOC}, there is just one value of { the terminal-price coefficient $\varphi_1$ for which $g(0)=0$. For all other $\varphi_1$, we have $g(0)\neq 0$ and an equilibrium exists.}

\item The equilibrium stock-price drift $\hat \mu_t$ and price levels $\hat{S}_t$ have the following qualitative properties: The second-order conditions  \eqref{SOC} ensure
\begin{align}\label{SOCCCC}
\begin{split}
&2(M+\overline{M})\kappa-(1+M+\overline{M})\nu_3>\kappa>0,\\
&2(M+\overline{M})\overline{\kappa}-(1+M+\overline{M})\bar{\nu}_3>\overline{\kappa}>0,
\end{split}
\end{align}
so that the equilibrium price-drift coefficients in \eqref{mu12} can be signed with $\mu_1(t) < 0$ and $\mu_2(t) < 0$. This is intuitive.  Larger latent aggregate trading-demand targets $\ta_\Sigma$ and larger hedging needs $\epsilon B_t$ mean that the equilibrium price drift $\hat \mu_t$ must be lower in order to incentivize price-sensitive targeted investors and hedgers to adjust their holdings to clear the market. Given negative price-drift coefficients $\mu_1(t)$ and $\mu_2(t)$, it then follows from \eqref{ODEs_equilibrium} that, given end-of-day terminal pricing coefficients $\varphi_0, \varphi_1 \ge0$, the intraday price-level coefficients $g(t)$ and $\zeta(t)$ are both positive.

\item The investor holding coefficients $\alpha_1(t), \ldots, \bar{\alpha}_2(t)$ in \eqref{optimaltheta_alphas} can also be signed using \eqref{mu12} and \eqref{SOCCCC}.  The targeted-investor coefficients $\alpha_1(t)$ and $\alpha_2(t)$ on the aggregate imbalance state variables $\ta_\Sigma$ and $\epsilon B_t$ are both negative.  This is intuitive because investors reduce their personal holdings in response to the lower price drifts induced in equilibrium by positive latent aggregate imbalances.  A similar intuition applies for the negative hedger coefficient $\bar\alpha_1(t)$ on the latent aggregate rebalancer imbalance $\ta_\Sigma$.  The coefficient $\alpha_3(t)$ on the targeted investor's personal target $\ta_i$ is, as expected, positive.   Similarly, the sign of the hedger's coefficient $\bar\alpha_2(t)$ on the hedging imbalance state variable is positive.  This is the net effect of $\epsilon B_t$ both as a personal target in the hedger penalty $L_{i,t}$ in \eqref{defLL} and as a state variable for the impact of the aggregate latent hedger imbalance on the price drift in the investor trading profits $X_{i,t}$ in \eqref{dX}. Recall that all initial stock positions have been normalized to  zero for all traders (with no loss of generality). Therefore, from \eqref{optimaltheta_ismall}, there are initial discrete orders (i.e., block trades $\theta_{i,0}-\theta_{i,-}\neq 0$) at $t=0$. This is related to why the initial price $\hat{S}_0$ fully reveals $\taS$. However, afterwards trading evolves continuously for $t\in(0,1]$.\footnote{The discontinuous model in Example \ref{ex:gamma} below has optimal discrete orders throughout the trading day.} 

\item Inserting $\bar{\nu}_0(t)$ from \eqref{nubar0} in Appendix \ref{app:proofs} into \eqref{dSmu8}  and rearranging using \eqref{optimaltheta_ismall} and \eqref{optimaltheta_alphas} for the equilibrium holdings $\hat{\theta}_{i,t}$ and \eqref{mu12} and \eqref{dSmuuuu} for the equilibrium price drift $\hat{\mu}_t$ lets us express the perceived price drift for an investor $i$ in \eqref{dSmu8} as
\begin{align}\label{eq:rearrangetheta}
\mu^{\theta_i}_t =
\begin{cases}
 \hat{\mu}_t + \tfrac{\nu_3(t) (M+\overline{M}+1)-2 \kappa(t)}{M+\overline{M}-1} \big(\theta_{i,t} - \hat \theta_{i,t}\big),\quad i\in\{1,...,M\},\\
 \hat{\mu}_t + \tfrac{\bar{\nu}_3(t) (M+\overline{M}+1)-2 \overline{\kappa}(t)}{M+\overline{M}-1} \big(\theta_{i,t} - \hat \theta_{i,t}\big),\quad i\in\{M+1,...,M+\overline{M}\} .
\end{cases}
\end{align}
This is not surprising given the equilibrium requirement that each investor perceives the same price drift in equilibrium, i.e., $\mu^{\hat{\theta}_i}_t = \hat{\mu}_t$ for all $i\in\{1,...,M+\overline{M}\}$.  Because the equilibrium holdings $\hat{\theta}_{i,t}$ include an $\ta_i$ term, the representation \eqref{eq:rearrangetheta} explains why there is an investor-specific $\ta_i$ term in the perceived price drift $\mu^{\theta_i}_t$ in \eqref{dSmu8}.

\item {The specific price-perception functions $\nu_0, \nu_1,$ and $\nu_2$ for the targeted investors are irrelevant in Theorem \ref{thm:Main} in the sense that, given $\nu_3$ and $\bar{\nu}_3$ and provided $\nu_0(t)\neq 0$, all different $\nu_0\neq 0,\,\nu_1$, and $\nu_2$ produce the same equilibrium prices and investor holdings. To provide some intuition, consider the drift in \eqref{dXj} for $j\in \{1,...,M+\bar{M}\}$. This drift is overparameterized because both $Y$ and $\theta_j$ are expressible in terms of $(\ta_j,\taS,B)$ when $Y$ is replaced by the solution $Y^{\theta_j}$ of \eqref{def:clearing_eq0}. We also note here that, while $\nu_0$, $\nu_1$, and $\nu_2$ do not affect equilibrium prices and investor holdings, they do pin down the other perceived price coefficients $\nu_4, \overline{\nu}_0, \overline{\nu}_1$, and $\overline{\nu}_2$ via \eqref{nubar0} and \eqref{nu4} in Appendix \ref{app:proofs}, which are related to market-clearing and the common perceptions of equilibrium price conditions from Definition \ref{def_eq}.

}

\end{enumerate}
\end{remark}

\bigskip

There is an important difference between prices in equilibrium and perceived { off-equilibrium} prices. Both in-equilibrium and off-equilibrium, the stock-price drifts $\hat \mu_t$ and $\mu^{\theta_i}_t$ are determined such that the market clears at each time $t$.  In the one case, this is part of the definition of equilibrium, and, in the other case, it is a reasonable off-equilibrium belief. However, in equilibrium, the price level $\hat S_t$ adjusts at time $t$ to be consistent with the required market-clearing drift $\hat \mu_t$ and the terminal price condition in \eqref{terminalS}. For example, a large target $\ta_\Sigma$ leads from \eqref{S_equilibrium} and \eqref{ODEs_equilibrium} to a high opening price $\hat S_0$ at $t=0$ so that the intraday drift $\hat \mu_t$ from  \eqref{mu12} and \eqref{dSmuuuu}   can be predictably low with prices drifting down in expectation over the day to the terminal price $\hat{S}_1$ in \eqref{terminalS}.  Similarly, a positive random shock to $B_t$ at time $t$ leads from \eqref{S_equilibrium} and \eqref{ODEs_equilibrium} to a random increase in prices $\hat S_t$ such that the market-clearing drift $\hat \mu_t$ can have a random decrease as required in  \eqref{mu12} and \eqref{dSmuuuu}. In contrast, for perceived off-equilibrium prices for investors with price-impact, we only specify how the perceived off-equilibrium price drifts $\mu^{\theta_i}_t$ in \eqref{dSmu8} change at each time $t$ so that the market clears given an investor's holdings $\theta_{i,t}$. However, the perceived off-equilibrium price level $S^{\theta_i}_t$ at time $t$ is not affected by $\theta_{i,t}$ at time $t$.  This simplifies the specification of off-equilibrium beliefs and is still reasonable for price beliefs since market-clearing depends on the price drift at time $t$, not on the contemporaneous price level (as per the discussion about the off-equilibrium price-impact model before \eqref{dSmu8}).

Price pressure $\hat S_t - D_t$ is the effect of intraday imbalances in latent trading demand. It can be positive or negative depending on the aggregate target imbalance $\ta_\Sigma$ and the hedging need $\epsilon B_t$. From \eqref{S_equilibrium},  price pressure has a deterministic trend $g(t) \ta_\Sigma$ over the day given the total rebalancing target imbalance $\ta_\Sigma$ and a stochastic component $\zeta(t) \epsilon B_t$ due to the randomly evolving hedging target.  For example, $g(0) \ta_\Sigma$ is the initial price impact of the aggregate latent trading-target imbalance $\ta_\Sigma$ revealed by the opening order-flow at time $t=0$.  Thereafter, the price impact $g(t) \ta_\Sigma$ of the imbalance $\ta_\Sigma$ varies predictably over time with $g(t)$.  A similar phenomenon applies to price pressure due to hedging demand. In particular, $\zeta(t) \epsilon dB_t$ is the immediate impact of an innovation $\epsilon dB_t$ in hedging demand at time $t$, and then $\zeta(s) \epsilon dB_t$ is the predictable continuation impact of $\epsilon dB_t$ at  times $s>t$ later in the day. From  \eqref{ODEs_equilibrium} and \eqref{mu12}, we see that $g(t)$ (the $\ta_\Sigma$  coefficient in $\hat S_t$) is affine in $\gamma(t)$, and that $\zeta(t) \epsilon$  (the $B_t$ coefficient in $\hat S_t$) is linear in $\epsilon$  (because $\zeta(t)$  does not depend on $(\gamma, \epsilon)$). Thus, the {path of intraday price pressure $ \hat{S}_t - D_t$ at different times $t$ during the day has an intertemporal factor-type} structure where $\ta_\Sigma$ is a common factor (which is different on different days but fixed over the course of a given day) that causes price pressure to change deterministically over the day given intraday variation in $g(t)$, and $B_t$ is a martingale that cause price pressure  {$ \hat{S}_t - D_t$} to change randomly over the day.  Thus, at a given time $s \in [0,1)$, the future random price pressure at subsequent times $t\in( s,1]$ have the following conditional means and variances:
\begin{align}\label{vol_liquidity_premium}
\begin{split}
\E[\hat S_t - D_t | \sigma(\ta_\Sigma, B_u)_{u\in[0,s]}] &= g(t) \ta_\Sigma + \zeta(t) \epsilon B_s,\\
\V[\hat S_t - D_t | \sigma(\ta_\Sigma, B_u)_{u\in[0,s]}] &= \zeta(t)^2 \epsilon^2 (t - s).
\end{split}
\end{align}
As a result, positive (negative) latent aggregate trading targets $\ta_\Sigma$ lead to predictable positive (negative) price-pressure trends $\E[\hat{S}_t |\sigma(\taS)]-D_t$ over the day.  In addition, randomness in the intraday hedging factor $B_t$ produces randomness in equilibrium price pressure, where higher hedging factors $B_t$ are associate with higher prices and lower price drifts. Again, this is intuitive.

\bigskip\noindent {\bf Empirical predictions:} Security prices are often decomposed econometrically into an informational component that follows a martingale and a residual liquidity effect (as in, e.g., Hasbrouck 1991).  A standard interpretation is that liquidity effects in prices decay predictably over time as liquidity supply from initial liquidity providers is first depleted by arriving order-flow imbalances and then replenished as order-flow imbalances are subsequently dispersed and absorbed by the broader market. Our model has two new empirical implications: First, our price pressure $\hat{S}_t - D_t$ is driven by both arriving orders and also by what those orders reveal about the underlying latent trading-demand imbalances $\taS$ and $\epsilon B_t$.  Second, the source of intraday predictability in our price pressure differs from the standard liquidity-supply interpretation. Price predictability here reflects the net effect of predictable time variation in latent trading demand (i.e., the targets $\gamma(t)\taS$ and $\epsilon B_t$) as well as predictability in liquidity supply (controlled by $\kappa(t)$ and $\bar{\kappa}(t)$). 

Empirical predictions from our model are testable using different types of data.  First, using standard intraday price and order-flow data (e.g., TAQ), the formula for equilibrium prices in \eqref{S_equilibrium} predicts intraday prices are driven by time-varying effects (controlled by $g(t)$) of a daily aggregate parent target that is constant over the day ($\ta_\Sigma$) plus a homoscedastic random walk ($D_t$) and a second Brownian motion ($B_t$) with heteroscedastic effects (controlled by $\zeta(t)\epsilon$).  Second, the price coefficients $g(t)$ and $\zeta(t)$ in 
\eqref{ODEs_equilibrium}  along with \eqref{mu12} give predictions for how the intraday effects ($g$ and $\zeta$ in the first prediction) vary across different days given daily variation in the  numbers of targeted rebalancers and hedgers ($M$ and $\overline{M}$) and given daily variation in order-execution costs due to variation in bid-ask spreads ({since order-execution costs are }implicitly represented by $\kappa$ and $\bar \kappa$). This can be estimated using individual-trader data (e.g., from IIROC) and standard data on bid-ask quotes. %Third, Remark \ref{rmkthm}.2 implies that the volatility of the {price pressure} is greater on days with more volatile order imbalances $\epsilon B_t$ due to a larger hedging coefficient $\epsilon$ and that intraday price price away from the fundamental price $D_t$ are stronger when $\gamma(t)$ is larger.

\section{Competitive equilibrium and welfare}\label{sec:Radner}
From Theorem \ref{thm:Main}, the deterministic functions $\nu_3(t)$ and $\bar{\nu}_3(t)$ for perceived own-order price-impacts are two degrees of freedom in our model. By imposing additional economically-motivated structure on $\nu_3(t)$ and $\bar{\nu}_3(t)$, we can identify unique equilibria corresponding to different forms of competition and market power. From Section \ref{sec:22}, price taking is a special case of interest.

In the competitive Radner equilibrium, investors act like price-takers over the whole day --- not just at the market open --- in that the perceived prices $S^{\theta_i}_t$ for each investor $i$ are unaffected by her holdings $\theta_{i,t}$. Hence, the competitive Radner equilibrium is obtained by requiring that the coefficient in front of $\theta_{i,t}$ in the perceived drift \eqref{dSmu8} for $dS^{\theta_i}_t$ is zero.  From \eqref{eq:rearrangetheta}, this requirement on $\nu_3(t)$ and $\bar{\nu}_3(t)$ is seen to imply
\begin{align}\label{Radner_zero_drift}
\begin{split}
\frac{\nu_3(t) (M+\overline{M}+1)-2 \kappa(t)}{M+\overline{M}-1}=0,\\
\frac{\bar{\nu}_3(t) (M+\overline{M}+1)-2 \overline{\kappa}(t)}{M+\overline{M}-1}=0,
%0&=\kappa  (-2 \overline{\kappa} \nu_0+2 \nu_0 \bar{\nu}_3+\nu_3 \overline{M} \bar{\nu}_0)-\nu_3 \big((M+1) \nu_0 (\bar{\nu}_3-\overline{\kappa})+\nu_3 \overline{M} \bar{\nu}_0\big),\\
%0&=2 \overline{\kappa} \bar{\nu}_0 (\nu_3-\kappa )+\overline{\kappa} M \nu_0 \bar{\nu}_3-M \nu_0 \bar{\nu}_3^2+(\overline{M}+1) \bar{\nu}_0 \bar{\nu}_3 (\kappa -\nu_3).
\end{split}
\end{align}
which gives the competitive-equilibrium perceived pricing coefficients\footnote{At first glance, it might seem that price-taking would mean price perceptions for investor $i$ with $\nu_3(t) = 0$ (for targeted investors) and $\bar \nu_3(t) = 0$ (for hedgers). However, investor $i$'s holdings $\theta_{i,t}$ have both a direct effect ($\nu_3$ and $\bar \nu_3)$ on perceived market-clearing prices in \eqref{Smu} and also an indirect effect via the optimal responses $\theta_{j,t}^{Y^{\theta_i}}$ of other investors $j \ne i$ to $\theta_{i,t}$ via the perceived market-clearing condition \eqref{def:clearing_eq0} and the endogenous perceived state-process $Y^{\theta_i}_t$ solving \eqref{def:clearing_eq0}. The conditions in  \eqref{Radner_zero_drift} ensure that investor holdings $\theta_{i,t}$ have no perceived price impact net of both effects.}
\begin{align}\label{max_nu}
\nu_3^*(t)=\frac{2 \kappa(t)}{1 + M + \overline{M}},\quad \bar{\nu}_3^*(t)=\frac{2 \overline{\kappa}(t)}{1 + M + \overline{M}}. 
\end{align}

We show next that the competitive Radner equilibrium is the welfare-maximizing equilibrium. While there are many ways to measure social welfare (see, e.g., Section 6.1 in Vayanos 1999), we follow Du and Zhu (2017, Eq. 42) and consider maximizing the expected aggregate certainty-equivalent for the $M+\overline{M}$ investors.  The certainty equivalent CE$_i\in\R$ for investor $i$ is defined by 
\begin{align}\label{CE_linear}
\begin{split}
\text{CE}_i &:= V_i(X_{i,0}), \quad i\in\{1,...,M+\overline{M}\},
\end{split}
\end{align}
where the value functions $V_i$ are defined in \eqref{objective}. The aggregate expected social welfare objective is given by
 \begin{align}\label{welfaremax0}
\begin{split}
\sup_{\nu_3(t),\bar{\nu}_3(t)}\sum_{i=1}^{M+\overline{M}}\E[\text{CE}_i],
%&\mu_3^\text{rebal}(t) \in \text{argmax}_{\mu_3(t)}\sum_{i=1}^{M}\E[\text{CE}_i],\\
%&\mu_3^\text{liq}(t) \in \text{argmax}_{\mu_3(t)}\sum_{i=M+1}^{M+\overline{M}}\E[\text{CE}_i],
\end{split}
\end{align}
subject to $(\nu_3,\bar{\nu}_3)$ satisfying requirements \eqref{SOC}. {The expectation in the objective \eqref{welfaremax0}  is ex ante in the sense that it is taken over random daily investor variables} $(\ta_1,...,\ta_M)$ and $(\theta_{1,-}^c,....,\theta_{M+\overline{M},-}^c)$. The following theorem shows that the competitive Radner equilibrium \eqref{max_nu} attains \eqref{welfaremax0}.

\begin{theorem}\label{welfaremax} In the setting of Theorem \ref{thm:Main}, let the random private targets $(\ta_1,...,\ta_M)$ be  square integrable and not perfectly correlated, and let $M\ge2$. 
The competitive Radner equilibrium with \eqref{max_nu} is the welfare-maximizing equilibrium in that the unique maximizers of \eqref{welfaremax0} are given by  {$\nu_3^*$ and $\bar{\nu}_3^*$ in} \eqref{max_nu}. The corresponding optimal holding strategies \eqref{optimaltheta_ismall} are given by
\begin{align}\label{social_welfare_thetas}
\hat{\theta}_{i,t}=
\begin{cases}
 \gamma(t) \ta_i- \frac{\overline{\kappa}(t)\big( \gamma(t) \ta_\Sigma+  \overline{M} \epsilon B_t\big) }{\overline{M} \kappa(t) + M \overline{\kappa}(t)}, \quad i =1,...,M,\\
 \frac{- \kappa(t)\gamma(t) \ta_\Sigma  + M \overline{\kappa}(t) \epsilon B_t }{\overline{M} \kappa + M \overline{\kappa}(t)}, \quad i=M+1,...,M+\overline{M},
 \end{cases}
\end{align}
and the corresponding equilibrium stock-price drift coefficients in \eqref{dSmuuuu} are given by 
\begin{align} \label{common_drift_Radner}
%\hat{\mu}_t&=  \mu_1(t)\gamma(t) \taS+\mu_2(t)B_t,\quad 
\mu_1(t):=- \frac{2\kappa(t) \overline{\kappa}(t)}{\overline{M} \kappa(t) + M \overline{\kappa}(t)},\quad \mu_2(t):=- \frac{2\kappa(t) \overline{\kappa}(t) \overline{M} \epsilon }{\overline{M} \kappa(t) + M \overline{\kappa}(t)}.
\end{align}
% When $M\sum_{i=1}^M \E[\ta_i^2]=\E[\ta_\Sigma^2]$, the set of maximizers for \eqref{social_welfare_max} is 
%\begin{align}
%\big\{ (\nu_3^*,\bar{\nu}_3^*):   \nu_3^*> \tfrac{2\kappa}{1+M+\overline{M}}-\tfrac{M(M+\overline{M}-1)\overline{\kappa}}{\overline{M}(1+M+\overline{M})}, \,\, 
%\bar{\nu}_3^*=\tfrac{2\overline{M}\kappa+2M\overline{\kappa}-\overline{M}(1+M+\overline{M})\nu_3^*}{M(1+M+\overline{M})}  \big\}.
%\end{align}
%In this case, $\nu_3^*$ is not unique and the drift-coefficient for $\theta_{i,t}$ in \eqref{dSmu8} of $dS^{\theta_i}_t$ may or may not be zero.
\end{theorem}

{ The proof (see Appendix \ref{app:proofs}) uses the non-trivial correlations between $(\ta_1,...,\ta_M)$ to produce a strict inequality in Cauchy-Schwartz's inequality. This ensures that the second-order condition for optimality corresponding to \eqref{welfaremax0} holds.}

A natural question concerns the impact of {benchmarks like TWAP and Almgren-Chriss targets} on competitive financial markets. By inserting \eqref{max_nu} into \eqref{mu12}, the price coefficients \eqref{ODEs_equilibrium} become
\begin{align}\label{zeta_RADNER}
\begin{split}
g(t)&=\varphi_1+2\int_t^1 \frac{ \kappa(u)  \overline{\kappa}(u)  }{\overline{\kappa}(u) M+\kappa(u) \overline{M}}\gamma(u)du,\\
\zeta(t) &=\varphi_0  + 2\overline{M}\int_t^1\frac{ \kappa(u)  \overline{\kappa}(u)  }{\overline{\kappa}(u) M+\kappa(u) \overline{M}}du.
\end{split}
\end{align}

\noindent From \eqref{zeta_RADNER}, when $\varphi_0>0$ and $\varphi_1>0$,  then both functions $g(t)$ and $\zeta(t)$ are positive, and equilibrium prices $\hat{S}_t$ differ from $D_t$. In the special case of $\varphi_0:=\varphi_1:=0$, the formulas in \eqref{zeta_RADNER} show that if either $\kappa$ or $\overline{\kappa}$ {is set to zero} (but not both) over $[0,1]$, then $g$ and $\zeta$ become 0, and, thus, $\hat{S}_t$ becomes $D_t$. This is because liquidity supply becomes infinite with $L_{i,1}=0$ for some investors. 

{ For $u\in[0,1)$, the integrand term $\frac{ \kappa(u)  \overline{\kappa}(u)  }{\overline{\kappa}(u) M+\kappa(u) \overline{M}}$ in \eqref{zeta_RADNER} is increasing in $\kappa(u)$ because its derivative with respect to $\kappa(u)$ is $\frac{M\overline{\kappa}(u)^2  }{(\overline{\kappa}(u) M+\kappa(u) \overline{M})^2}\ge0$. Therefore, given $M, \overline{M} \ge 1$, the functions $g(t)$ and $\zeta(t)$ in \eqref{zeta_RADNER} are increasing in $\kappa$  in the sense that $\kappa(u) \le \kappa^\circ(u)$ for $u\in[0,t]$   produces corresponding ordered solutions $g(t)\le g^\circ(t)$ and $\zeta(t)\le \zeta^\circ(t)$.}  Hence, when the incremental order-execution costs and inventory and risk-management penalties for trading deviations from the target trajectory $\gamma(t) \ta_i$ are large, then predictable trends in intraday price pressure $g(t) \ta_\Sigma$ are large, and market liquidity $\zeta(t)$ for hedging trading imbalances $\epsilon B_t$ is low, and  {price-pressure variance $\zeta(t)^2\epsilon^2 t$} is high. This leads to the following comparison result:

\begin{corollary}\label{cor1} In the competitive Radner equilibrium with $\nu^*_3$ and $\bar{\nu}_3^*$ from \eqref{max_nu}, the function $\zeta(t)$ in \eqref{zeta_RADNER} is increasing in $\kappa(t)$ and $\bar{\kappa}(t)$. Consequently, for a fixed function $\overline{\kappa}(t)$ and a time point $t^\circ\in(0,1)$, for two penalty-severity functions $\kappa(t)$ and $\kappa^\circ(t)$ ordered such that $\kappa(t) < \kappa^\circ(t)$ for $t \in [0, t^\circ)$ and $\kappa(t) = \kappa^\circ(t)$ for $t \in [t^\circ,1]$, the  illiquidity $\zeta(t)$ and price volatility in \eqref{vol_liquidity_premium} are less in the market with $\kappa(t)$ than in the market with $\kappa^\circ(t)$.  The same is true for analogous $\bar{\kappa}(t)$ and $\bar{\kappa}^\circ(t)$.

\end{corollary}

\proof The first claim follows from $\zeta$'s representation in \eqref{zeta_RADNER}. The second claim follows from the representation of $\hat{S}_t-D_t$ in \eqref{S_equilibrium}. The third claim follows from symmetry.

$\endproof$

\noindent In addition, in the competitive equilibrium, predictable trends in intraday price pressure $g(t)\taS$ are increasing in the target ratio $\gamma(t)$, and price volatility is increasing in the hedging {scalar} $\epsilon$. 

For markets to clear, the equilibrium holdings $\hat{\theta}_{i,t}$ of targeted investors and hedgers differ from their targets.  In the competitive equilibrium, these differences, from \eqref{social_welfare_thetas}, are 
\begin{align}\label{social_welfare_thetas2}
\begin{split}
\hat{\theta}_{i,t}- \gamma(t) \ta_i&=
- \frac{\gamma(t) \ta_\Sigma+  \overline{M} \epsilon B_t}{\overline{M} \frac{\kappa(t)}{\overline{\kappa}(t)} + M}, \quad i =1,...,M,\\
\hat{\theta}_{i,t}-\epsilon B_t&= - \frac{ \gamma(t) \ta_\Sigma+  \overline{M} \epsilon B_t }{\overline{M} + M \frac{\overline{\kappa}(t)}{\kappa(t)}}, \quad i=M+1,...,M+\overline{M}.
\end{split}
\end{align}
Thus, competitive rebalancers and liquidity providers split the aggregate imbalances {$ \gamma(t)\taS + \overline{M} \epsilon B_t$} with the hedgers independently of their individual targets $\ta_i$. The sharing is pro rata adjusted for their differential penalty severities. { In this context, we note that intraday liquidity providers (i.e., targeted investors with targets $\ta_i = 0$) absorb a share of the aggregate demand imbalance, but not the full imbalance given their inventory-holding penalties from \eqref{defL}. Our results are} consistent with evidence in van Kerval, Kwan, and Westerholm (2018) that dynamic trading by large investors involves reciprocal liquidity provision (if investor targets are in opposite directions {and cancel out in aggregate}) or reduced trading (if their targets are in the same direction). The limits of the differences in \eqref{social_welfare_thetas2} as time $t\uparrow 1$ depend on the limiting behavior of the relative penalty severities $\frac{\overline{\kappa}(t)}{\kappa(t)}$. Section \ref{sec:numerics} gives examples of $\kappa(t)$ and $\overline{\kappa}(t)$ and illustrates different possible limits of \eqref{social_welfare_thetas2}.

In non-competitive equilibria, investors behave strategically over $t\in(0,1]$ in that they perceive their holdings have an impact on prices.\footnote{Even with non-price-taking behavior over $t\in (0,1]$, our model still requires price-taking at the market open $t = 0$ such that \eqref{S0} holds for the opening price $S_0$ both on- and off-equilibrium.  Allowing for non-price-taking behavior at $t = 0$ would complicate the measurability condition in \eqref{S0}, and, thus, might require investors to filter over time to estimate $\ta_\Sigma$ rather than being able to infer it from $S_0$.  Modeling targeted trading in this more complicated learning environment would be an interesting future extension.} In particular, the $\theta_{i,t}$-coefficients in the perceived drift \eqref{dSmu8} for $dS^{\theta_i}_t$ are non-zero.  A natural non-competitive specification is that these drift coefficients would be negative so that increased positive holdings $\theta_{i,t}$ would depress perceived price drifts.  Once again, the individual non-competitive strategies aggregate such that the resulting equilibrium prices only depend on the aggregate state variables $\ta_\Sigma$ and $\epsilon B_t$.  However, if {investors are not price-takers and }\eqref{max_nu} does not hold, then the individual rebalancer targets $\ta_i$ can appear in the rebalancer deviation $\hat{\theta}_{i,t}- \gamma(t) \ta_i$ in addition to $\taS$ and $\epsilon B_t$. \ \\

\noindent{\bf Empirical predictions:} Eq.  \eqref{zeta_RADNER} and Corollary \ref{cor1} lead to a set of empirical predictions about changing conditions across different days in a competitive market. Suppose that on different days there are different numbers of  investors $M$ (rebalancers and liquidity providers) and $\overline{M}$ (hedgers) with different penalty severities $\kappa$ and $\overline{\kappa}$. Our model predicts the intraday price-pressure variance $ \zeta(t)^2 \epsilon^2 t$ and market illiquidity $\zeta(t)$ should be higher on days on which there are fewer rebalancers and liquidity providers $M$ and larger penalty severities $\kappa$ and $\overline{\kappa}$. These results are not normative critiques of TWAP trading but rather positive predictions about the empirical effect of targeted trading on daily market dynamics.

\section{Equilibria with stochastic targets}\label{sec:VWAP} 

Because our traders have linear preferences, they behave myopically in the sense that the solution to their individual optimization problems \eqref{objective} is found by maximizing pointwise at each time $t\in[0,1]$. Therefore, Theorem \ref{thm:Main}(i) continues to hold for equilibrium investor holdings $\hat{\theta}_{i,t}$ when the deterministic target ratio function $\gamma(t)$ is replaced with an arbitrary stochastic process $\gamma = (\gamma_t)_{t\in[0,1]}$ that is independent of the Brownian motions $(D,B)$ and the private targets $(\ta_1,...,\ta_M)$. A natural interpretation is that random intraday fluctuations in implicit  bid-ask order-execution costs lead to changes in the target ratio $\gamma_t$.  Once again, the rebalancer penalty process $L_{i,t}$ is a reduced-form for incremental order-execution, {inventory},  and risk-management costs relative to the {now stochastic} target trajectory $\gamma_t \ta_i$. Provided $\gamma_t$ is observable for all investors at time $t\in[0,1]$, we can re-define the filtrations in  \eqref{meas1} as
\begin{align}\label{Fit_gamma}
 \sF_{i,t}&:=
\begin{cases}
\sigma(S_u,D_u,B_u,\gamma_u,\tilde{a}_i,\theta^c_{i,-})_{u\in[0,t]}, \quad i=1,...,M,\\
\sigma(S_u,D_u,B_u,\gamma_u,\theta^c_{i,-})_{u\in[0,t]}, \quad i=M+1,...,M+\overline{M}.
\end{cases}
\end{align} 

Lemma \ref{lem:respone}  in Appendix \ref{app:proofs} (trader $j$'s optimal response)  continues to hold word-for-word when $\gamma(t)$ is replaced with $\gamma_t$ and the martingale $N_t$ in \eqref{dNt} is replaced by either of the two martingales in \eqref{mg_BM} and \eqref{mg_gamma} below. From the proof of Lemma \ref{lem:respone}, the reason  this extension is possible is that traders with linear utilities solve for their optimal holdings via pointwise optimization {at each $t\in[0,1]$}. 

Theorem \ref{thm:Main}(ii) for equilibrium prices $\hat{S}_t$ needs to be adjusted depending on the specification of the stochastic dynamics of the target-ratio process $\gamma_t$. We consider two specific examples of  stochastic target ratios. Both examples have a zero initial value $\gamma_0=0$, and both tie down the terminal target ratio by requiring $\gamma_1 =1$ at the end of the day.

\begin{example}[Brownian bridge]\label{ex:BMbridge} In this target specification, the deterministic target ratio $\gamma(t)$ is replaced with the stochastic target-ratio process $\gamma_t$ defined as the Brownian bridge process solving the stochastic differential equation
\begin{align}\label{BM_bridge1}
d\gamma_t := \frac{1-\gamma_t}{1-t}dt + dZ_t,\quad \gamma_0:=0,
\end{align}
where $Z=(Z_t)_{t\in[0,1]}$ is an independent standard Brownian motion. At time $s$, the conditional expected future target ratio at time $t>s$ is
\begin{align}
\E_s[\gamma_t] = \gamma_s + \frac{t-s}{1-s}(1-\gamma_s),\quad 0\le s< t< 1.
\end{align}
Moreover, the drift in \eqref{BM_bridge1} ensures that $\gamma_t \to 1$ almost surely as $t\uparrow 1$. To derive the appropriate version of Theorem \ref{thm:Main}(ii), we redefine the candidate stock price \eqref{S_equilibrium} as
\begin{align}\label{S_VWAPBM}
\hat{S}_t :=D_t+\big( h(t)+ \sigma(t)\gamma_t\big)\taS+ \zeta(t)\epsilon B_t,\quad t\in[0,1],
\end{align}
for three deterministic functions $h(t),\sigma(t)$, and $\zeta(t)$ solving the { system of linear  ODEs:}
\begin{align}\label{ODEs_equilibrium_BM}
\begin{split}
\sigma'(t) &=\frac{\sigma(t)}{1-t}+ \mu_1(t),\quad  \sigma(1)=0,\\
h'(t) &=-\frac{\sigma(t)}{1-t},\quad  h(1)= \varphi_1,\\
\zeta'(t) &=\mu_2(t),\quad  \zeta(1)= \varphi_0,
\end{split}
\end{align}
where the deterministic functions $\mu_1$ and $\mu_2$ are unchanged from \eqref{mu12}. { The linear ODE system \eqref{ODEs_equilibrium_BM} is triangular in the following sense: First, the ODE for $\sigma(t)$ in \eqref{ODEs_equilibrium_BM} is explicitly solved in \eqref{dSmuuuuBM} below. Second, given $\sigma(t)$, the remaining two ODEs in \eqref{ODEs_equilibrium_BM} for $h(t)$ and $\zeta(t)$ are solved by integrating, where their solutions are given in \eqref{ODEs_equilibrium_BMg} and \eqref{ODEs_equilibrium}}.   Furthermore, the martingale $N_t$ in \eqref{S_equilibriumm} is redefined as follows:
\begin{align}\label{mg_BM}
dN_t := dD_t +\zeta(t) \epsilon dB_t+ \sigma(t)\taS dZ_t,\quad N_0:=0.
\end{align}
Thus, a qualitatively new feature with a stochastic target ratio is that intraday price pressure due to the target imbalance $\taS$ is now random due to the target-ratio shocks $dZ_t$ in \eqref{BM_bridge1}.

\begin{theorem}[{Brownian bridge}]\label{thm:Main_BM} Under the assumptions of Theorem \ref{thm:Main}, $h(0)\neq0$, {and when the target ratio $\gamma_t$ is a Brownian bridge}, there exists an equilibrium in which:
\begin{itemize}
\item[(i)] The perceived price parameters $\nu_0,...,\nu_4$ and $\bar{\nu}_0,...,\bar{\nu}_3$ are as in Theorem \ref{thm:Main}.

\item[(ii)] Investor equilibrium holdings are given by 
\begin{align} \label{optimaltheta_ismall_BM} 
\hat{\theta}_{i,t} = 
\begin{cases}
 \alpha_1(t)\gamma_t\taS+\alpha_2(t)\epsilon B_t+\alpha_3(t) \gamma_t\ta_i,\quad i\in\{1,...,M\},\\
\bar{\alpha}_1(t)\gamma_t\taS+\bar{\alpha}_2(t)\epsilon B_t,\quad i\in\{M+1,...,M+\overline{M}\}, 
\end{cases}
\end{align}
where the deterministic functions $\alpha_1,\alpha_2,\alpha_3,\bar{\alpha}_1$, and $\bar{\alpha}_2$ are given by \eqref{optimaltheta_ismall}.
\item[(iii)] The equilibrium stock price is defined by \eqref{S_VWAPBM} with the martingale $N_t$ given by \eqref{mg_BM} for deterministic functions $h,\sigma$, and $\zeta$ that are the unique solutions of the linear ODE system in \eqref{ODEs_equilibrium_BM}, and the price drift  is given by
\begin{align} \label{dSmuuuuBM} 
\hat{\mu}_t &:= \mu_1(t)\gamma_t\taS+\mu_2(t)\epsilon B_t,
\end{align}
where the deterministic functions $\mu_1$ and $\mu_2$ are given by \eqref{mu12}. Furthermore, the linear ODE for $\sigma$ in \eqref{ODEs_equilibrium_BM} is uniquely solved by 
\begin{align}
\sigma(t) &= -\frac1{1-t} \int_t^1 (1-u) \mu_1(u)du\label{sigmaSgamma},\quad t\in[0,1),
\end{align}
which satisfies $\lim_{t\uparrow 1}\sigma(t) =0$. The solution \eqref{sigmaSgamma} ensures that $\tfrac{\sigma(t)}{1-t}$ is integrable; hence, $h(t)$ in \eqref{ODEs_equilibrium_BM} is found by integration for $t\in[0,1]$:
\begin{align}\label{ODEs_equilibrium_BMg}
\begin{split}
h(t) &=\varphi_1+\int_t^1 \frac{\sigma(u)}{(1-u)}du.
%\zeta(t) &=\varphi_0-\int_t^1 \mu_2(u)du.
\end{split}
\end{align}
The solution for $\zeta(t)$ is identical to \eqref{ODEs_equilibrium}.
\end{itemize}
\end{theorem}
$\endproof$
\end{example}

The {coefficient $h(t) + \sigma(t) \gamma_t$ giving the price impact }of the target imbalance $\ta_\Sigma$ with a Brownian bridge target ratio $\gamma_t$ in \eqref{S_VWAPBM}  is related to the corresponding coefficient $g(t)$ with a deterministic target ratio $\gamma(t)$ as follows:  Consider a deterministic target ratio equal to the ex ante expected Brownian bridge target ratio $ \gamma(t):=\E[\gamma_t]=t$ and let $g(t)$ denote the associated deterministic price impact in \eqref{ODEs_equilibrium}.  Then, for $t\in[0,1]$:
\begin{align}\label{eq:bridgederiv}
\begin{split}
\E[h(t) + \sigma(t) \gamma_t] 
&= h(t) + \sigma(t) \gamma(t) \\
&=g(t),
\end{split}
\end{align}
where the last equality holds because at $t=1$ we have $g(1) =\varphi_1 = h(1)$ and for $t<1$ the time-derivatives of both sides of \eqref{eq:bridgederiv} agree.\footnote{For $t\in(0,1)$, Eq. \eqref{ODEs_equilibrium} produces $g'(t)=\mu_1(t)\gamma(t)=\mu_1(t)t$ whereas \eqref{sigmaSgamma} and \eqref{ODEs_equilibrium_BMg}  produce $\sigma'(t)=\frac{\sigma(t)}{1-t} +\mu_1(t)$ and $h'(t)=-\frac{\sigma(t)}{1-t}$. Combining these with the product rule produces the claim. }
In other words, the price impact $h(t) + \sigma(t) \gamma_t$ in \eqref{S_VWAPBM} with a stochastic { target ratio} $\gamma_t$ has the same ex ante expected price impact as in the corresponding deterministic model with $\gamma(t)=\E[\gamma_t]$ plus additional randomness.

An unrealistic feature of the Brownian bridge target ratio is that $\gamma_t$ can be negative as well as bigger than one with positive probability at times $t\in(0,1)$. Our next target-ratio process does not have these problems. The following construction is based on gamma processes.\footnote{Recall that a L\'evy process $l=(l_t)_{t\in[0,1]}$ with $l_0:=0$ and gamma distributed increments $l_t-l_s$,  $0\le s <t\le 1$, is called a gamma process. In our case, the mean and variance are normalized to one.  A gamma bridge process is then defined by $\gamma_t:=\frac{l_t}{l_1}$ for $t\in[0,1]$. See \'Emery and Yor (2004) as well as Frei and Westray (2015) for more details.} Such pure jump processes have a long history of applications in option pricing theory (see, e.g., Madan, Carr, and Chang 1998). 

%\begin{align}\label{psis}
%\psi_0(t):= \int_{[0,1]} (1-z)^{-t}dz=\tfrac1{1-t},\quad \psi_1(t):= \int_{[0,1]} z(1-z)^{-t}dz=\tfrac{1}{(1-t)(2-t)},
%\end{align}
%for $t\in[0,1)$.

\begin{example}[Gamma bridge]\label{ex:gamma} The following model is based on Frei and Westray (2015). In this model variation, the deterministic target ratio $\gamma(t)$ is replaced with a stochastic target-ratio  process $\gamma_t$ 
that is a c\`adl\`ag gamma bridge process starting at $\gamma_0=0$ and ending at $\gamma_1=1$. In between the gamma bridge increases via a series of positive jumps that are dense on $(0,1]$. Corollary 1 in \'Emery and Yor (2004) ensures that $\gamma_t$ has predictable intensity $\frac{1-\gamma_{t-}}{1-t}$ and that the quadratic variation process $[\gamma]_t$ has predictable intensity $\frac{(1-\gamma_{t-})^2}{(1-t)(2-t)}$ where $\gamma_{t-}:= \lim_{s\uparrow t} \gamma_t$ for $t\in(0,1]$ is the left-continuous version of the c\`adl\`ag process $\gamma_t$. In other words, the compensated processes
\begin{align}\label{bridge_intensity}
\gamma_t - \int_0^t \frac{1-\gamma_{s-}}{1-s}ds,\quad\text{and}\quad  [\gamma]_t-\int_0^t \frac{(1-\gamma_{s-})^2}{(1-s)(2-s)}ds,\quad t\in[0,1],
\end{align} 
are martingales.\footnote{For simplicity, the underlying gamma process is normalized to have unit mean and unit variance, which among other properties gives us $\E[\gamma_t] =t$ and $\E[\gamma^2_t] =\tfrac12t(1+t)$. However, the following analysis can easily be modified to include a parameter $m\in(0,\infty)$ by redefining the predictable intensity of the quadratic variation process $[\gamma]_t$ to be $\frac{(1-\gamma_{t-})^2}{(1-t)(1+m(1-t))}$ and leaving $\gamma$'s predictable intensity as $\frac{1-\gamma_{t-}}{1-t}$. This would give us $\E[\gamma^2_t] =\tfrac{t(1+mt)}{1+m}$ whereas $\E[\gamma_t] =t$ is as before.} 

The equilibrium holdings $\hat{\theta}_{i,t}$ in the gamma bridge model are unchanged from \eqref{optimaltheta_ismall} in Theorem \ref{thm:Main}(i).
To derive the appropriate version of Theorem \ref{thm:Main}(ii) for the equilibrium stock price $\hat{S}_t$, the candidate stock price \eqref{S_equilibrium} is again redefined by \eqref{S_VWAPBM} for deterministic functions $(h,\sigma,\zeta)$, and the martingale $N_t$ is redefined as 
\begin{align}\label{mg_gamma}
dN_t := dD_t +\zeta(t)\epsilon dB_t+\sigma(t)\taS \big(d\gamma_t - \tfrac{1-\gamma_{t-}}{1-t}dt\big),\quad N_0:=0.
\end{align}

Consider now the requirements in Definition \ref{def_admis} in this setting. First, the quadratic variation process $[N]_t$ of $N_t$ has dynamics given by
\begin{align}
d[ N]_t = \big(1+\zeta(t)^2\epsilon^2\big)dt +  \sigma(t)^2\taS^2 d[\gamma]_t.
\end{align}
Consequently, the second martingale in \eqref{bridge_intensity} produces dynamics for the predictable quadratic variation process (i.e., $[N]_t$'s compensator; see, e.g.,  p.122 in Protter 2004) as 
\begin{align}
d\langle N \rangle_t= \Big(1+\zeta(t)^2\epsilon^2+ \sigma(t)^2  \taS^2\frac{(1-\gamma_{t-})^2}{(1-t)(2-t)}\Big)dt.
\end{align}
{Second, while holdings $\theta_{i,t}$ are} required to be adapted to the filtration $\sF_{i,t}$ defined in \eqref{Fit_gamma}, the left-continuity requirement in Definition \ref{def_admis} (part of the c\`agl\`ad requirement on admissible $\theta_{i,t}$) prevents trader $i\in \{1,...,M\}$ from using holdings that depend on $\gamma_t$ such as, e.g., $\ta_i\gamma_t$. This is because gamma bridges are not left-continuous. However, $\theta_{i,t}$ can depend on the left-continuous version $\gamma_{t-}$.\footnote{This admissibility restriction is essentially an assumption about how quickly investors can act on $\gamma_t$.}

\begin{theorem}[{ Gamma bridge}]\label{thm:Main_gamma} Under the assumptions of Theorem \ref{thm:Main}, $h(0)\neq0$,  and when the target ratio $\gamma_t$ is a gamma bridge, there exists an equilibrium in which:
\begin{itemize}
\item[(i)] The perceived price parameters $\nu_0,...,\nu_4$ and $\bar{\nu}_0,...,\bar{\nu}_3$ are as in Theorem \ref{thm:Main}.

\item[(ii)] Investor equilibrium holdings are given by 
\begin{align} \label{optimaltheta_ismall_gamma}
\hat{\theta}_{i,t} = 
\begin{cases}
 \alpha_1(t)\taS\gamma_{t-}+\alpha_2(t)\epsilon B_t+\alpha_3(t)\ta_i \gamma_{t-},\quad i\in\{1,...,M\}, \\
 \bar{\alpha}_1(t)\taS\gamma_{t-}+\bar{\alpha}_2(t)\epsilon B_t,\quad i\in\{M+1,...,M+\overline{M}\},
\end{cases}
\end{align}
where the deterministic functions $\alpha_1,\alpha_2,\alpha_3,\bar{\alpha}_1$, and $\bar{\alpha}_2$ are given by \eqref{optimaltheta_ismall}.
\item[(iii)] The equilibrium stock price is defined by \eqref{S_VWAPBM} with the martingale $N_t$ defined in \eqref{mg_gamma} for deterministic functions $h,\sigma$, and $\zeta$ given as the unique solutions of the linear ODE system in \eqref{ODEs_equilibrium_BM}, and the price drift  is given by
\begin{align} \label{dSmuuuugamma} 
\hat{\mu}_t &:= \mu_1(t)\taS\gamma_{t-}+ \mu_2(t)\epsilon B_t,
\end{align}
where the deterministic functions $\mu_1$ and $\mu_2$ are again defined by \eqref{mu12}.
\end{itemize}
\end{theorem}

$\endproof$

\end{example}

The equilibria with the { Brownian bridge and gamma bridge } target-ratio processes $\gamma_t$ in Theorem \ref{thm:Main_BM}  and Theorem \ref{thm:Main_gamma} both have the following comparative statistics:

\begin{corollary}\label{cor2} In the setting of Theorem \ref{thm:Main_BM} and Theorem \ref{thm:Main_gamma}, the dynamics of the predictable quadratic variation process of the equilibrium  { price pressure} $\hat{S}_t-D_t$ and  the predictable quadratic cross-variation processes between $\hat{S}_t-D_t$ and $(\gamma,B)$ are given by  
\begin{align}\label{VWAPcrossvar}
\begin{split}
d\langle \hat{S}-D\rangle_t &= \sigma(t)^2\taS ^2d\langle \gamma\rangle_t+\zeta(t)^2\epsilon ^2dt,\\
d\langle \hat{S}-D,\gamma\rangle_t &=\sigma(t)\taS d\langle \gamma\rangle_t,\\
\quad d\langle \hat{S}-D,B\rangle_t &=  \zeta(t)\epsilon dt.
\end{split}
\end{align}
In the Brownian bridge model \eqref{BM_bridge1} we have $d\langle \gamma\rangle_t =dt$  whereas in the gamma bridge model the second martingale in \eqref{bridge_intensity} produces $d\langle \gamma\rangle_t=\tfrac{(1-\gamma_{t-})^2}{(1-t)(2-t)}dt$.
\end{corollary}

\proof  {From Theorem \ref{thm:Main_BM} and Theorem \ref{thm:Main_gamma} the equilibrium stock-price process is \eqref{S_VWAPBM} for both the Brownian and gamma bridge processes.} The variations \eqref{VWAPcrossvar} follow from the representation of $\hat{S}_t-D_t$ from \eqref{S_VWAPBM}.

$\endproof$

\noindent The formulas in \eqref{VWAPcrossvar} show that comovement between  price pressure and the underlying sources of randomness are completely determined by the solutions of the ODEs in \eqref{ODEs_equilibrium_BM}. Consequently, variances and covariances for the price pressure $\hat{S}_t-D_t$ can be expressed in terms of these functions. For example, in the gamma bridge model, the price-pressure variance is
\begin{align}
\begin{split}
\V[\hat{S}_t-D_t|\sigma(\taS)] &=\E\big[\big( \sigma(t)(\gamma_t-t)\taS+ \zeta(t)\epsilon B_t\big)^2|\sigma(\taS)\big] \\
&=  \sigma(s)^2\E[(\gamma_t-t)^2]\taS^2+ \zeta(t)^2 \epsilon^2 t\\
&=  \sigma(s)^2\frac{t(1-t)}2\taS^2+ \zeta(t)^2 \epsilon^2 t,
\end{split}
\end{align}
where the last equality uses $\E[\gamma_t]=t$ and $\E[\gamma_t^2] = \frac12t(1+t)$.
Section \ref{sec:numerics} provides numerical examples of the solutions to the ODEs in \eqref{ODEs_equilibrium_BM}. \ \\

\noindent{\bf Empirical predictions:} In the competitive Radner equilibrium where $\nu^*_3$ and $\bar{\nu}_3^*$ are given by \eqref{max_nu}, the representation of $\zeta$ in \eqref{zeta_RADNER} continues to hold whereas $\sigma$ in \eqref{sigmaSgamma} becomes 
\begin{align}\label{sigma_RADNER}
\sigma(t) &= \frac1{1-t} \int_t^1 \frac{2(1-u)\kappa(u) \overline{\kappa}(u)}{\overline{M} \kappa(u) + M \overline{\kappa}(u)}du,\quad t\in[0,1).
\end{align}
Consequently, in the competitive Radner equilibrium, the comparative statistics for price volatility given in Corollary \ref{cor1} continue to hold (i.e., price volatility is increasing in $\kappa$).

\section{Connection to VWAP benchmark trading}\label{connVWAP}
The stochastic target ratio $\gamma_t$ in Section \ref{sec:VWAP}  is driven by randomly evolving market conditions that affect intraday order-execution costs. This section shows how the stochastic target ratio $\gamma_t$ can be linked to VWAP trading. To do this, we augment our earlier model so that aggregate market volume is now the sum of two components: First, as before, there is trading by the rebalancers, liquidity providers, and hedgers. Second, there is now additional trading by a new group of investors that we introduce into the model.  These are a large number of other buyers and sellers --- who we call the \emph{crowd} --- with inelastic trading demands that are assumed to naturally net to zero. For example, these could be retail investors and small asset managers trading for naturally offsetting personal reasons.  In our augmented VWAP model, these two groups trade alongside each other. Since trading by the crowd is naturally offsetting, it has no impact on aggregate trading-demand imbalances. However, it does contribute to the aggregate market volume. We assume here that crowd volume affects (or is correlated with) order-execution costs and, thus, affects the trading target trajectories $\gamma_t \ta_i$ of the targeted investors via $\gamma_t$.  In contrast, trading by our targeted investors affects aggregate order imbalances and, thus, market-clearing prices. However, we assume there is no feedback loop whereby their trading affects order-execution costs and, thus, their (exogenous in our model) intraday trading target trajectories.\footnote { In practice, VWAP benchmarking for an investor often excludes the investor's own trading from the measure of volume used contractually in VWAP benchmarking (see, e.g., Madhavan 2002, Exhibit 1). For simplicity, we keep $\epsilon$ constant for hedgers. Allowing for a stochastic $\epsilon_t$ is also possible. } 

To this end, let $\text{vol} = (\text{vol}_t)_{t\in[0,1]}$  be an exogenous stochastic process for the stock's cumulative crowd volume { vol$_t$ over the interval $[0,t]$ for  times $t\in[0,1]$.} The relative cumulative volume process is then defined as the ratio
\begin{align}\label{VWAP1} 
 \frac{\text{vol}_t}{\text{vol}_1},\quad t\in[0,1],
\end{align}
which is zero initially, has non-decreasing paths, and has terminal value one. Because vol$_1$ at time $t=1$ cannot be observed at times $t < 1$, the volume ratio \eqref{VWAP1}  also cannot be observed at times $t < 1$. Consequently, the ratio \eqref{VWAP1} cannot be used as a state-process. 

The VWAP objective replacing \eqref{objective} for targeted investors is
\begin{align}\label{VWAP2}
\sup_{\theta_i\in \sA_i} \E\left[ X_{i,1} - \int_0^1 \kappa(t) \Big( \frac{\text{vol}_t}{\text{vol}_1}\ta_i - \theta_{i,t} \Big)^2dt\,\Big|\, \sF_{i,0}\right],\quad i\in\{1,...,M\}.
\end{align}
The idea behind \eqref{VWAP2} is that intraday fluctuations in the relative volume ratio \eqref{VWAP1} affect the ex post intraday trading target trajectory $\frac{\text{vol}_t}{\text{vol}_1}\ta_i$.  However, because investor $i$'s position $\theta_{i,t}$ is adapted to the filtration ${\cal F}_{i,t}$ and because investor $i$ cannot use her holdings $\theta_{i,t}$ to manipulate the intraday volume weights $\frac{\text{vol}_t}{\text{vol}_1}$,  the optimization problem \eqref{VWAP2} can be replaced, given linear utilities, with the equivalent problem:\footnote{Even though  \eqref{VWAP2} and  \eqref{VWAP3} yield different objective values, they are equivalent in the sense that they share the same maximizer. {This is because the terms in \eqref{VWAP2} and \eqref{VWAP3} involving $\theta_{i,t}$ are identical by iterated expectations. }} 
\begin{align}\label{VWAP3}
\sup_{\theta_i\in \sA_i} \E\left[ X_{i,1} - \int_0^1 \kappa(t) \Big(  \E\left[ \frac{\text{vol}_t}{\text{vol}_1}\Big|\sF_{i,t}\right] \ta_i -\theta_{i,t}\Big)^2dt\,\Big|\, \sF_{i,0}\right].
\end{align}
We model $\E\left[ \frac{\text{vol}_t}{\text{vol}_1}\Big|\sF_{i,t}\right]$ directly  as a gamma bridge $\gamma_t$. In that case,  \eqref{VWAP3}
becomes \eqref{objective} when $\sF_{i,t}$ is defined by \eqref{Fit_gamma} and $\gamma(t)$ is replaced by $\gamma_t$. Frei and Westray (2015) model  the realized relative volume curve $\frac{\text{vol}_t}{\text{vol}_1}$ used for VWAP benchmarking by the gamma bridge $\gamma_t$ from Example \ref{ex:gamma}. However, as discussed on page 617 in Frei and Westray (2015), this presents a potential problem because the realized relative volume curve in \eqref{VWAP1} cannot be observed prior to the end of the trading day. In contrast, the intraday expected relative volume curve $\E\left[\frac{\text{vol}_t}{\text{vol}_1}|\sF_{i,t}\right]$ in \eqref{VWAP3} in our model is --- by definition --- observable at times $t\in[0,1]$. {Thus,  we model it as a gamma bridge.}\ \\

\noindent {\bf Empirical implications:} Prices in the gamma bridge model in Section \ref{sec:VWAP}  include a stochastic response to random changes in the target ratio $\gamma_t$, which includes the effect of changing order-execution and inventory costs on the optimal target trajectory for targeted investors.  Our VWAP analysis links these fluctuations in the stochastic target ratio $\gamma_t$ to intraday fluctuations in the expected intraday volume ratio $\E\left[ \frac{\text{vol}_t}{\text{vol}_1}\big|\sF_{i,t}\right]$.  Thus, our VWAP model predicts that a positive shock  to the expected daily volume ratio at time $t$ increases the magnitude $|\hat{S}_t - D_t|$ of price pressure due to increased trading-demand imbalances due to an increase in the magnitude of the aggregate intraday trading target $|\gamma_t \ta_\Sigma|$.

\section{Numerics}\label{sec:numerics}
This section presents numerics for the competitive Radner equilibrium where \eqref{Radner_zero_drift} holds. Our analysis uses the two bridge models for the target ratio $\gamma_t$ in Examples \ref{ex:BMbridge} and \ref{ex:gamma}. The objects of interest are: (i) properties of the equilibrium price process $\hat{S}_t$ in \eqref{S_VWAPBM}, and (ii) how the rebalancers, liquidity providers, and hedgers optimally trade. The numerical properties here illustrate the analytic derivations in Section \ref{sec:VWAP}.

Our analysis uses the terminal stock-price restriction \eqref{S_1} (i.e., $\varphi_0:=\varphi_1:=0$ in \eqref{terminalS}), $M:= 10$ targeted investors, and $\overline{M}:=10$ hedgers. The private targets $(\ta_1,...,\ta_M)$ here are independent and have ex ante moments
\begin{align}\label{num0}
 \E[\ta_i]=0,\quad \E[\ta_i^2]=1,\quad i\in\{1,...,M\}.
\end{align}

\begin{figure}[!h]
\begin{center}
\caption{Examples of penalty severity functions $\kappa(t)$ in \eqref{num2}. The lines are: $p:=0$ (------), $\;p:=0.1\;(- - ),\;p:=0.5\;(-\cdot-),\;p:=0.99\;(-\cdot\cdot-).$}\ \\
$\begin{array}{c}
\includegraphics[width=7cm, height=5cm]{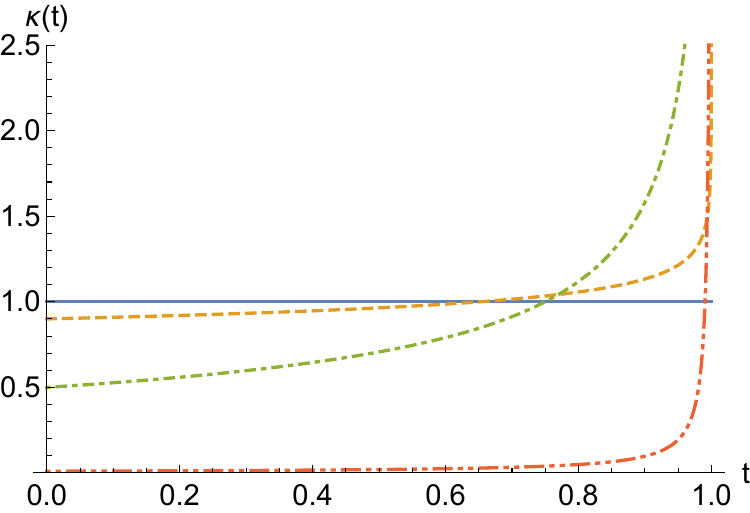} 
\end{array}$
\label{fig:kappa}
\end{center}
\end{figure}

We consider penalty-severity functions in \eqref{defL} and \eqref{defLL} defined by
\begin{align}\label{num2}
\begin{split}
\kappa(t) &:= \frac{(1-p)}{(1-t)^p},\quad t\in[0,1),\quad p\in[0,1),\\
\overline{\kappa}(t) &:= \frac{(1-\bar{p})}{(1-t)^{\bar{p}}},\quad t\in[0,1),\quad \bar{p}\in [0,1),
%\kappa_3(t)&:=
%\begin{cases} .01\quad \text{for}\; t\in[0,0.85], \\
%-38.0132 + 44.7214t \quad \text{for}\;  t\in[0.85,0.95],\\ 
% \frac{1}{\sqrt{1-t}}\quad  \text{for}\;  t\in[0.95,1). 
 %\end{cases}
\end{split}
\end{align}
which are parameterized by $p$ and $\bar{p}$. For comparison purposes, these functions both integrate to one over $t\in[0,1]$. A natural baseline is $p:=0$, where the penalty severity is constant over the day.  For $p\in(0,1)$, the penalty functions explode as $t \uparrow 1$ at various rates but still satisfy the integrability condition \eqref{square_integrable}.   When $p$ is close to one (e.g., $p:=0.99$), the intraday penalty severities are negligible (i.e., close to zero) until close to the end of the day. In this case, our model mimics the situation where trader $i\in\{1,...,M\}$ faces no intraday penalties but just a quasi-hard terminal constraint $\theta_{i,1}=\ta_i$. Fig. \ref{fig:kappa} illustrates some of the $\kappa(t)$ functions used in our numerical analysis.

\begin{figure}[!h]
\begin{center}
\caption{Equilibrium price functions $\sigma(t), h(t)$ and $\zeta(t)$ in \eqref{sigma_RADNER}, \eqref{ODEs_equilibrium_BMg}, and \eqref{zeta_RADNER} for the competitive equilibrium \eqref{max_nu}. The parameters are given by \eqref{num0}-\eqref{num2}, $\epsilon:=1$, and the discretization divides the day $t\in [0,1]$ into 1,000 trading rounds. The lines are $p=\bar{p}=0$ (------), $p=0.5,\,\bar{p}=0\; (- - ),\,p=0,\,\bar{p}=0.5\;(-\cdot-),\;p=\bar{p}=0.5\;(-\cdot\cdot-).$}\ \\
$\begin{array}{cc}
\includegraphics[width=7cm, height=5cm]{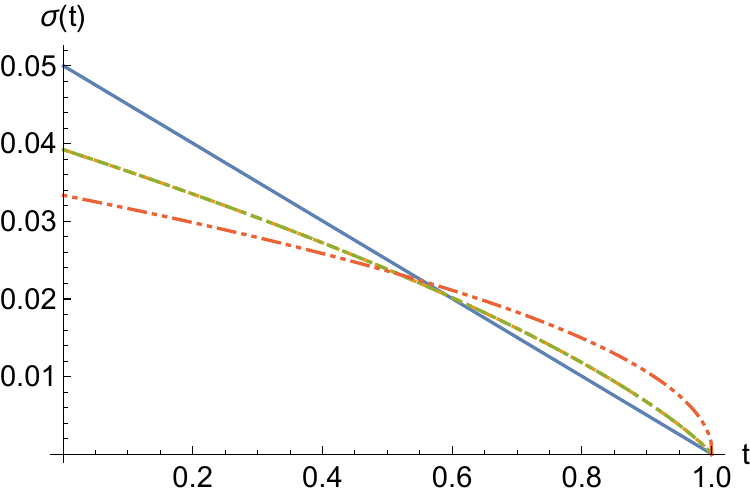} &\includegraphics[width=7cm, height=5cm]{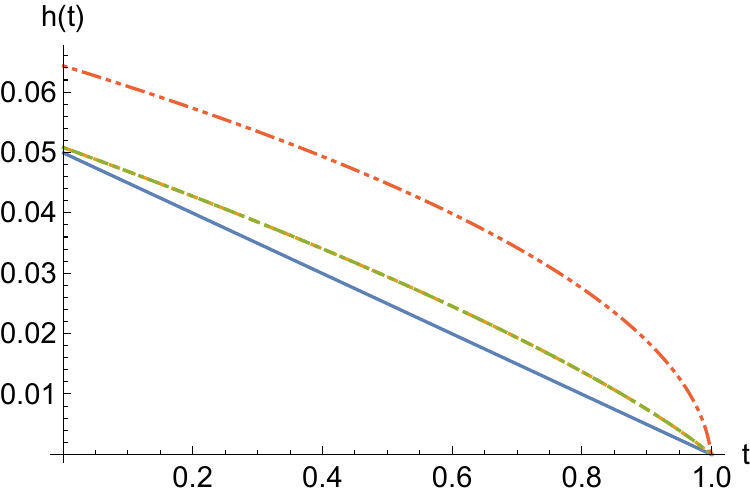}\\
\includegraphics[width=7cm, height=5cm]{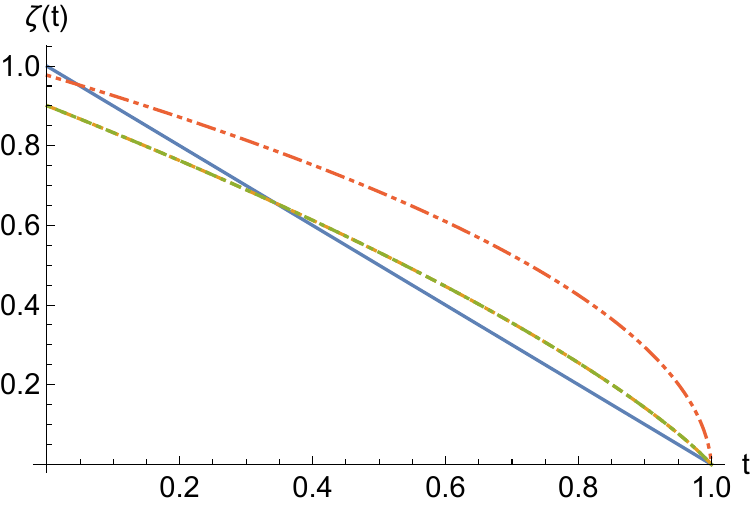} &
%\\ \text{A: [Welfare] } & \text{B: [Radner]}
\end{array}$
$\begin{array}{c}
\end{array}$
\label{fig:sigma}
\end{center}
\end{figure}

Our first topic is equilibrium pricing. From Theorems \ref{thm:Main}, \ref{thm:Main_BM}, and \ref{thm:Main_gamma}, the price function $\zeta(t)$ in \eqref{S_equilibrium} and \eqref{S_VWAPBM} is identical for the deterministic $\gamma(t)$ model and for both the Brownian and gamma bridge $\gamma_t$ models and is given in \eqref{zeta_RADNER}. The price-loading functions $\sigma(t)$ and $h(t)$ do not appear in the deterministic $\gamma(t)$ model but are the same for both the Brownian and gamma bridge models and are given in \eqref{sigma_RADNER} and \eqref{ODEs_equilibrium_BMg}. Fig. \ref{fig:sigma} shows the price-loading functions $\sigma(t)$, $h(t)$, and $\zeta(t)$ for different values of $p$ and $\bar{p}$. The signs of $\sigma(t)$, $h(t)$, and $\zeta(t)$ are all positive (from \eqref{sigmaSgamma}, \eqref{ODEs_equilibrium_BMg},  and Remark \ref{rmkthm}.2). We note that the greater the penalty severity $\kappa(t)$ is, the more sensitive prices are to shocks in the amount $\epsilon B_t$ driving the hedger trading demand, which the targeted investors must provide. The values $\zeta(t)$ and $h(t)$ converge to $\varphi_0$ and $\varphi_1$ as $t\uparrow 1$, which in these numerics are, {for simplicity, } taken to be zero.

\begin{figure}[!h]
\begin{center}
\caption{$\taS$-coefficients in the conditional expected deviations \eqref{Edeviation} and \eqref{HFTdeviation} for the competitive equilibrium \eqref{max_nu}. The parameters are given by \eqref{num0}-\eqref{num2}, $\epsilon:=1$, and the discretization divides  the day $t\in [0,1]$ into  1,000 trading rounds. The lines are: $p=\bar{p}=0$ (------), $p=0.5,\,\bar{p}=0\; (- - ),\,p=0,\,\bar{p}=0.5\;(-\cdot-),\;p=\bar{p}=0.5\;(-\cdot\cdot-).$}\ \\
$\begin{array}{cc}
\includegraphics[width=7cm, height=5cm]{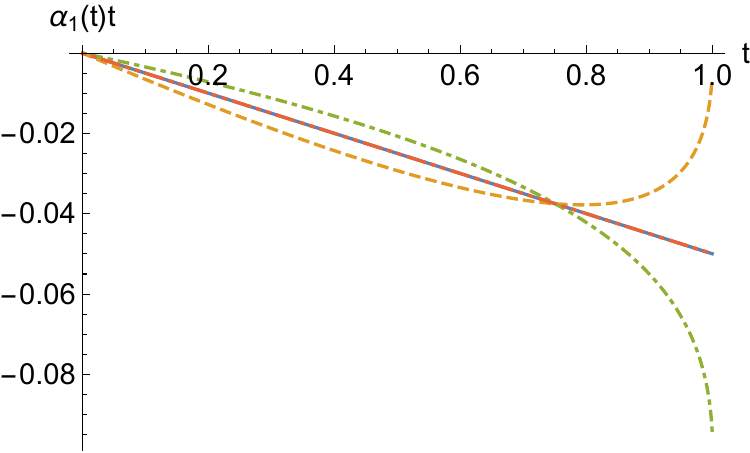} &\includegraphics[width=7cm, height=5cm]{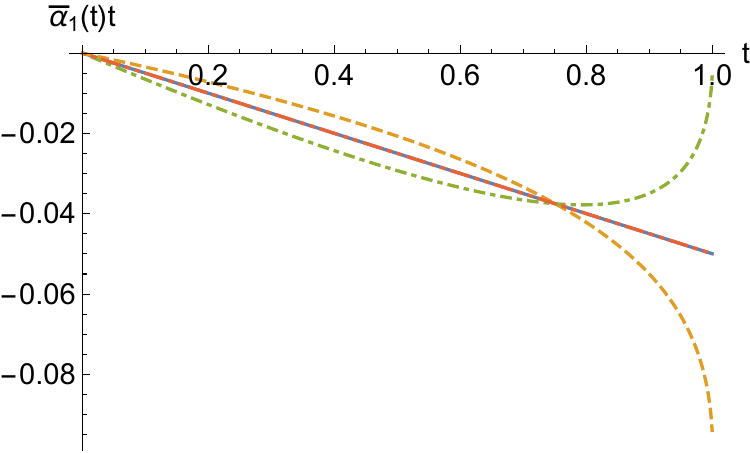}
\end{array}$
\label{fig:Edeviation}
\end{center}
\end{figure}

Our second topic is the equilibrium stock holdings. For $i\in\{1,...,M\}$, the optimal VWAP strategies for $i\in\{1,...,M\}$ in \eqref{optimaltheta_ismall_BM} and \eqref{optimaltheta_ismall_gamma} give expected holdings over the day
\begin{align}\label{eq:TWAP-deviation} 
\begin{split}
\E[\hat{\theta}_{i,t}|\sigma(\ta_i,\taS)] &= \alpha_1(t) \E[\gamma_{t-}]\taS+\alpha_2(t) \epsilon\E[B_t] + \alpha_3(t)\E[\gamma_{t-}] \ta_i\\
&= \alpha_1(t) t \taS+ \alpha_3(t) t \ta_i.
\end{split}
\end{align}
Consequently, from \eqref{eq:TWAP-deviation}, trader $i\in\{1,...,M\}$ expects ex ante to deviate from her target trajectory $\ta_i\gamma_{t-}$ by
\begin{align}\label{Edeviation}
\begin{split}
\E[\hat{\theta}_{i,t}-  \ta_i\gamma_{t-} |\sigma(\ta_i,\taS)]&=\alpha_1(t) t \taS+ \big(\alpha_3(t) -1\big)t \ta_i\\
&=\alpha_1(t) t \taS,
\end{split}
\end{align}
where the last equality follows from inserting the competitive values \eqref{max_nu} into the expression for $\alpha_3$ given in \eqref{optimaltheta_alphas}{, which gives $\alpha_3(t)=1$ here}. Fig. \ref{fig:Edeviation} shows the coefficient on $\taS$ in the expected deviations between targeted investor $i$'s holdings up through time $t\in[0,1]$ relative to her corresponding target.  Similarly, for hedger $i\in\{M+1,...,M+\overline{M}\}$, the optimal strategies \eqref{optimaltheta_ismall_BM} and \eqref{optimaltheta_ismall_gamma} give ex ante expected hedger target deviations
\begin{align}\label{HFTdeviation}
\begin{split}
\E[\hat{\theta}_{i,t}-\epsilon B_t|\sigma(\taS)]&= \bar{\alpha}_1(t) t \taS.
\end{split}
\end{align}

Finally, we illustrate that when $p\neq \bar{p}$, the targeted investors' and hedgers' soft target constraints can become hard constraints. In other words, if the targeted investor penalty severity $\kappa(t)$ explodes faster as $t\uparrow 1$ than the hedger penalty severity $\overline{\kappa}(t)$, the targeted investors hit their targets $\ta_i$ with probability one at the end of trading. To see this, we use \eqref{social_welfare_thetas2}  to compute the limit. Because  $\gamma_1=1$, the terminal deviation for targeted investors $i\in \{1,...,M\}$ is
\begin{align}\label{differences_example}
\begin{split}
\hat{\theta}_{i,1}- \ta_i&=\begin{cases}
0, & \bar{p}<p,\\
-\frac{\ta_\Sigma + \overline{M} \epsilon B_1}{\overline{M}+M}, & \bar{p}=p,\\
-\frac{ \ta_\Sigma + \overline{M} \epsilon B_1}{M}, & \bar{p}>p.
\end{cases}
\end{split}
\end{align}
For hedgers $i\in \{M+1,...,M+\overline{M}\}$, the terminal deviation is
\begin{align}\label{differences_example2}
\begin{split}
\hat{\theta}_{i,1}-\epsilon B_1 &=\begin{cases}
-\frac{ \ta_\Sigma + \overline{M} \epsilon B_1 }{\overline{M}}, & \bar{p}<p,\\
-\frac{\ta_\Sigma + \overline{M} \epsilon B_1 }{\overline{M}+M}, & \bar{p}=p,\\
0& \bar{p}>p.
\end{cases}
\end{split}
\end{align}

{
\section{Extension to inhomogeneous investors}
We can allow for inhomogeneity within the realtime hedgers. To illustrate this, we split the group of real-time hedgers into two homogenous subgroups with respectively $\bar{M}$ and $M^\circ$ investors. Therefore, we replace \eqref{defLL} with hedger penalties
\begin{align}\label{defL2}
L_{i,t}:=
\begin{cases}
\int_0^t \overline{\kappa}(s)\big(\theta_{i,s}-\epsilon B_s\big)^2ds,\quad i\in\{M+1,...,M+\bar{M}\},\\
\int_0^t \kappa^\circ(s)\big(\theta_{i,s}-\epsilon^\circ B_s\big)^2ds,\quad i\in\{M+\bar{M}+1,...,M+\bar{M}+M^\circ\},
\end{cases}
\end{align} 
for $ t\in[0,1]$. In the penalty processes in \eqref{defL2}, the two subgroups' penalty-severity functions $\kappa^\circ(t)$ and $\bar{\kappa}(t)$ can differ and their scalars $\epsilon^\circ$ and $\epsilon$ can differ.

Up to minor modifications, the equilibrium structure in Theorem \ref{thm:Main} continues 
hold, but for this model extension there are now three free perceived price-impact parameters $(\nu_3,\bar{\nu}_3,\nu_3^{\circ})$. The 
equilibrium stock-price drift in \eqref{S_equilibriumm} becomes 
\begin{align}
\begin{split}
\hat{\mu}_t& = -\frac{(M+\bM+\cM)C \bC \cC (2\kappa+\nu_4)}{2\big((M+\bM+\cM)^2-1\big)(\kappa-\nu_3)(\cM C \bC+\bM C \cC+M \bC \cC)} \gamma \tilde{a}_\Sigma\\
&\quad - \frac{2 C ( \bM \bk \cC \epsilon + \cM \ck \bC \ce)}{\cM C \bC+\bM C \cC+M \bC \cC} B_t,
\end{split}
\end{align}
where 
\begin{align}
\begin{split}
C(t)&:=2(M+\bM+\cM)\kappa(t)-(1+M+\bM+\cM)\nu_3(t),\\
\bar{C}(t)&:=2(M+\bM+\cM)\bk(t)-(1+M+\bM+\cM)\bar{\nu}_3(t),\\
\cC(t)&:=2(M+\bM+\cM)\ck(t)-(1+M+\bM+\cM)\nu_3^{\circ} (t),\\
\nu_4&:=-\frac{4\kappa^2+2(1+M+\bM+\cM)(-2+M+\bM+\cM)\kappa \nu_3}{(M+\bM+\cM)C}.
\end{split}
\end{align}
Likewise, the optimal investor holdings in \eqref{optimaltheta_ismall} become \begin{align}
\begin{split}
\hat{\theta}_t&= -\frac{(M+\bM+\cM)\bC \cC(2\kappa+\nu_4)}{2(1+M+\bM+\cM)(\kappa-\nu_3)(\cM C \bC+\bM C \cC+M \bC \cC)} \gamma \tilde{a}_\Sigma\\
&\quad - \frac{2(-1+M+\bM+\cM)(\bM \bk \cC \epsilon + \cM \ck \bC \ce)}{\cM C \bC+\bM C \cC+M \bC \cC} B_t  \\
&\quad + \frac{(M+\bM+\cM)(2\kappa+\nu_4)}{2(1+M+\bM+\cM)(\kappa-\nu_3)} \gamma \tilde{a}_i,\\
\hat{\bar{\theta}}_t&= -\frac{(M+\bM+\cM)C \cC(2\kappa+\nu_4)}{2(1+M+\bM+\cM)(\kappa-\nu_3)(\cM C \bC+\bM C \cC+M \bC \cC)} \gamma \tilde{a}_\Sigma\\
&\quad + \frac{2(-1+M+\bM+\cM)(\bk(M \cC+\cM C) \epsilon - \cM \ck C \ce)}{\cM C \bC+\bM C \cC+M \bC \cC} B_t,  \\
\hat{\theta}^\circ_t&= -\frac{(M+\bM+\cM)C \bC(2\kappa+\nu_4)}{2(1+M+\bM+\cM)(\kappa-\nu_3)(\cM C \bC+\bM C \cC+M \bC \cC)} \gamma \tilde{a}_\Sigma\\
&\quad + \frac{2(-1+M+\bM+\cM)(\ck(M \bC+\bM C) \ce - \bM \bk C \epsilon)}{\cM C \bC+\bM C \cC+M \bC \cC} B_t.
\end{split}
\end{align}

We conjecture that the model structure scales linearly with as many free perceived hedger price-impact coefficients $\bar{\nu}_3$ as there are heterogenous hedger subgroups.
 }

\section{ Discussion}

Two features of the mathematical structure of our model are particularly important for tractability. The first is that the aggregate target $\ta_\Sigma$ is inferable at time $t=0$ and so, as a result, there is no need for investors to filter trading data over time to estimate the other investors' private targets.  The second is the linear preference structure,  which leads to the optimal controls $\hat{\theta}_{i,t}$ at time $t\in[0,1]$ being solutions to pointwise optimization problems.

There are good reasons to think the qualitative properties of our analysis are robust even though the specific functional forms of prices and trading strategies depend on our modeling assumptions (e.g., Brownian motion dynamics).  First, investor perceptions about how prices respond to off-equilibrium orders are likely a key factor determining the equilibrium form.  Second, the absence of manipulative predatory trading is still likely with rational forward-looking liquidity provision. Third, intraday liquidity is likely to be reduced by intraday trading target penalties relative to just terminal end-of-day target penalties. 

Lastly, we comment on numerical implementation. The model is characterized by low dimensional state-processes that makes numerics fast to perform. Furthermore, the model's linear structure makes the numerics stable (coupled linear ODEs). We have experimented extensively with the numerics and have not found any instability concerns.

\section{Conclusion}

This paper has solved for continuous-time { Subgame Perfect }Nash equilibria with endogenous liquidity provision and intraday trading targets. We show how TWAP, VWAP, and other trading benchmarks induce intraday patterns in investor positions and in price dynamics. There are also potential extensions of our model.  First, it would be interesting to extend the model to allow for more heterogeneity in the  investor optimization problems. For example, $\gamma(t)$ and $\epsilon$ could be replaced  with different ratios $\gamma_i(t)$ for targeted investors and $\epsilon_i$ or a stochastic process $\epsilon_t$ for hedgers. Second, perhaps the most pertinent extension would be to allow for additional randomness such that the initial equilibrium stock price $\hat{S}_0$ cannot fully reveal the aggregate target $\taS$.  Such an extension would require filtering to learn about trading-demand imbalances.  {Third, the Brownian motion driving hedger demands might be private information rather than publicly observable.}

\appendix

\section{Proofs}\label{app:proofs}

\begin{lemma}[Trader $j$'s optimal response]\label{lem:respone}  Let $\nu_0(t)\neq0$, $\bar{\nu}_0(t)\neq0$, and assume that \eqref{SOC} holds. Fix an exogenous state-process with c\`agl\`ad paths $Y = (Y_t)_{t\in[0,1]}$ and fix a trader index  $j\in \{1,...,M+\overline{M}\}$. When the stock price $S$ in  the wealth \eqref{dX} and filtration $\sF_{i,t}$ in \eqref{meas1} is  $S:=S^Y$ with drift \eqref{dXj}
 and martingale \eqref{dNt}, the optimizer for \eqref{objective} over $\sA_j$ is 
\begin{align}\label{NASHH1}
\theta^{Y}_{j,t} &:=
\begin{footnotesize}
\begin{cases}
\frac{1}{2\big(\kappa(t)-\nu_3(t)\big)}\Big(\nu_0(t)Y_{t} +\nu_1(t)\gamma(t)\taS+\nu_2(t)B_t +\big(\nu_4(t)+2\kappa(t)\big)\gamma(t)\ta_j\Big),\quad j \in \{1,...,M\},\\
\frac{1}{2\big(\overline{\kappa}(t)-\bar{\nu}_3(t)\big)}\Big(\bar{\nu}_0(t)Y_{t} +\bar{\nu}_1(t)\gamma(t)\taS+\big(\bar{\nu}_2(t)+2\epsilon \overline{\kappa}(t)\big)B_t\Big), \quad j \in \{M+1,...,M+\overline{M}\},
\end{cases}
\end{footnotesize}
\end{align}
provided that $\theta^{Y}_{j,t}$ satisfies the integrability condition \eqref{admis_integrability}.
\end{lemma}

\noindent \emph{Proof of Lemma \ref{lem:respone}}: Because utilities are linear, we have the following representation for an arbitrary control $\theta_j \in \sA_j$ for targeted investors $j\in\{1,...,M\}$
\begin{align}\label{resp1}
\begin{split}
&\E[X_{j,1} - L_{j,1}|\sF_{j,0}] 
\\&=X_{j,0} + \E\Big[ \int_0^1\Big(\theta_{j,s}\big(\nu_0(s) Y_s +\nu_1(s)\gamma(s)\taS+\nu_2(s)B_s+\nu_3(s)\theta_{j,s}+ \nu_4(s)\gamma(s)\ta_j \big) \\&\quad \quad \quad \quad \quad - \kappa(s)\big(\theta_{j,s}-\gamma(s)\tilde{a}_j\big)^2\Big)ds\,\Big|\,\sF_{j,0}\Big].
\end{split}
\end{align}
For hedgers $j\in\{M+1,...,M+\overline{M}\}$, we have a similar representation
\begin{align}\label{resp2}
\begin{split}
&\E[X_{j,1} - L_{j,1}|\sF_{j,0}] \\
&=X_{j,0} + \E\Big[ \int_0^1\Big(\theta_{j,s}\big(\bar{\nu}_{0}(s)Y_s +\bar{\nu}_1(s)\gamma(s)\taS+ \bar{\nu}_2(s)B_s+\bar{\nu}_3(s)\theta_{j,s}\big) \\&\quad \quad \quad \quad \quad - \overline{\kappa}(s)\big(\theta_{j,s}-\epsilon B_s\big)^2\Big)ds\,\Big|\,\sF_{j,0}\Big].
\end{split}
\end{align}
The martingale property of $\int\theta_j dN$ is used to eliminate $\E[\int_0^1 \theta_{j,s}dN_s]$ in both \eqref{resp1} and \eqref{resp2}.
The integrands in the $ds$-integrals are quadratic in $\theta_{j,s}$. Consequently, the supremum in \eqref{objective} is achieved by maximizing the integrands in \eqref{resp1} and \eqref{resp2} pointwise over $\theta_{j,s}$ at each state and at each time $s\in[0,1]$. This gives the optimal holdings in \eqref{NASHH1}. 

The left-continuity of $Y$'s paths allows investor $j$ to infer $Y$ from past and current observations of $S^Y$ and $(D,B)$. To see this, it suffices to show that observing 
\begin{align}\label{mu_i_infer}
\int_0^t \nu_0(s)Y_sds\text{ for } j \in\{1,...,M\} \text{ and } \int_0^t \bar{\nu}_0(s)Y_sds \text{ for }j\in\{M+1,...,M+\overline{M}\},
\end{align}
over time $t\in[0,1)$ is sufficient for investor $j$ to infer $Y_t$. The integrals in \eqref{mu_i_infer} are well-defined because $ \nu_0(s)$ and $\bar{\nu}_0(s)$ are continuous on $[0,1)$ and $Y$ has c\`agl\`ad paths (hence, $Y$'s paths are locally bounded). The ability to infer $Y_t$ from \eqref{mu_i_infer} follows directly from computing the time derivative from the left of \eqref{mu_i_infer}. This left-derivative is $\nu_0(t)Y_t$ for $j \in\{1,...,M\}$ and $\bar{\nu}_0(t)Y_t$ for $j\in\{M+1,...,M+\overline{M}\}$ by the left-continuity requirement placed on $Y$'s paths. Consequently, $Y_t$ is $\sF_{i,t}$ measurable when $S:= S^Y$ in \eqref{meas1} where $S^Y$ is defined using \eqref{dXj}. Therefore, $Y_t$ can be used as a state-variable for investor $j$ in \eqref{NASHH1}.

$\endproof$

\noindent {\bf Comment:} Because our traders are penalized for deviations {of their holdings} from intraday targets, investor optimal stock holdings are given in terms of levels rather than trading rates. This property allows our investors to absorb trading noise with only finite quadratic variation such as the Brownian motion dynamics in, e.g.,  Kyle (1985).

\proof[Proof of Lemma \ref{lem:new}] We define the state-process
\begin{footnotesize}
\begin{align}\label{def:clearing_eq01}
Y^{\theta_i}_t:=
\begin{cases}
\frac{2  \kappa  (\bar{\nu}_3-\overline{\kappa})-(\overline{\kappa}-\bar{\nu}_3) \big((M-1) \nu_1+\nu_4\big)+\overline{M} \bar{\nu}_1 (\nu_3-\kappa )}{(M-1) \nu_0 (\overline{\kappa}-\bar{\nu}_3)+\overline{M} \bar{\nu}_0 (\kappa -\nu_3)}\gamma\taS-\frac{(M-1) \nu_2 (\overline{\kappa}-\bar{\nu}_3)+\overline{M} (\kappa -\nu_3) (\bar{\nu}_2+2 \overline{\kappa} \epsilon )}{(M-1) \nu_0 (\overline{\kappa}-\bar{\nu}_3)+\overline{M} \bar{\nu}_0 (\kappa -\nu_3)}B_t
\\
\quad -\frac{2 (\kappa -\nu_3) (\overline{\kappa}-\bar{\nu}_3)}{(M-1) \nu_0 (\overline{\kappa}-\bar{\nu}_3)+\overline{M} \bar{\nu}_0 (\kappa -\nu_3)}\theta_{i,t} 
+\frac{(\overline{\kappa}-\bar{\nu}_3) (2 \kappa +\nu_4)}{(M-1) \nu_0 (\overline{\kappa}-\bar{\nu}_3)+\overline{M} \bar{\nu}_0 (\kappa -\nu_3)}\gamma \ta_i, \quad i\in\{1,...,M\},
\vspace{0.1cm}
\\
-\frac{2 \kappa  (\overline{\kappa}-\bar{\nu}_3)+(\overline{\kappa}-\bar{\nu}_3) (M \nu_1+\nu_4)+(\overline{M}-1) \bar{\nu}_1 (\kappa -\nu_3)}{M \nu_0 (\overline{\kappa}-\bar{\nu}_3)+(\overline{M}-1) \bar{\nu}_0 (\kappa -\nu_3)}\gamma \taS +\frac{M \nu_2 (\bar{\nu}_3-\overline{\kappa})-(\overline{M}-1) \bar{\nu}_2 (\kappa -\nu_3)-2 \overline{\kappa} (\overline{M}-1) \epsilon  (\kappa -\nu_3)}{M \nu_0 (\overline{\kappa}-\bar{\nu}_3)+(\overline{M}-1) \bar{\nu}_0 (\kappa -\nu_3)}B_t
\\
\quad -\frac{2 (\kappa -\nu_3) (\overline{\kappa}-\bar{\nu}_3)}{M \nu_0 (\overline{\kappa}-\bar{\nu}_3)+(\overline{M}-1) \bar{\nu}_0 (\kappa -\nu_3)}\theta_{i,t}, \quad i\in\{M+1,...,M+\overline{M}\},
\end{cases}
\end{align}
\end{footnotesize}
so that the corresponding drift process  \eqref{dXj} perceived by investor $j\neq i$ becomes
\begin{align}\label{dSmu}
\begin{split}
\begin{cases}
 \nu_0(t)Y^{\theta_i}_t +\nu_1(t)\gamma(t)\taS+\nu_2(t)B_t+\nu_3(t)\theta_{j,t}+ \nu_4(t)\gamma(t)\ta_j,\quad j\in\{1,...,M\},\\
 \bar{\nu}_0(t)Y^{\theta_i}_t +\bar{\nu}_1(t)\gamma(t)\taS+ \bar{\nu}_2(t)B_t+\bar{\nu}_3(t)\theta_{j,t},\quad j\in\{1+M,...,M+\overline{M}\}.
\end{cases}
\end{split}
\end{align}
Thus, trader $j$'s optimal response holdings $\theta^{Y^{\theta_i}}_{j,t}$  to $Y^{\theta_i}_t$ are given by \eqref{NASHH1} in Lemma \ref{lem:respone}. By summing these holdings, we see that \eqref{def:clearing_eq0} holds.

$\endproof$

\noindent \emph{Proof of Theorem \ref{thm:Main}}: We first insert $S^{\theta_i}$ from \eqref{Smu} into the objective \eqref{objective} to get
\begin{align}\label{appA1}
\begin{split}
&\E\Big[\int_0^1\Big(\theta_{i,t}\mu^{\theta_i}_t - \kappa(t)\big(\theta_{i,t}-\gamma(t)\tilde{a}_i\big)^2\Big)dt\Big],\quad i\in\{1,...,M\},\\
&\E\Big[\int_0^1\Big(\theta_{i,t}\mu^{\theta_i}_t - \overline{\kappa}(t)\big(\theta_{i,t}-\epsilon B_t\big)^2\Big)dt\Big],\quad i\in\{M+1,...,M+\overline{M}\},
\end{split}
\end{align}
where $\mu^{\theta_i}_t$ is defined in \eqref{dSmu8}. In \eqref{appA1}, the expectation $\E[\int_0^1 \theta_{i,t} dN_t]$ drops out because $\theta_i\in \sA_i$ ensures that the stochastic  integrals $\int \theta_{i} dN$ are martingales. The integrands in \eqref{appA1} are quadratic functions of $\theta_{i,t}$ and the quadratic terms produce the second-order conditions
\begin{align}\label{appAA2}
\begin{split}
&\frac{(\kappa -\nu_3) ((M+1) \nu_0 (\overline{\kappa}-\bar{\nu}_3)+\overline{M} \bar{\nu}_0 (\kappa -\nu_3))}{(M-1) \nu_0 (\overline{\kappa}-\bar{\nu}_3)+\overline{M} \bar{\nu}_0 (\kappa -\nu_3)}>0, \\
&\frac{(\overline{\kappa}-\bar{\nu}_3) (M \nu_0 (\overline{\kappa}-\bar{\nu}_3)+(\overline{M}+1) \bar{\nu}_0 (\kappa -\nu_3))}{M \nu_0 (\overline{\kappa}-\bar{\nu}_3)+(\overline{M}-1) \bar{\nu}_0 (\kappa -\nu_3)}>0.
\end{split}
\end{align}
Provided \eqref{appAA2} holds, the pointwise maximizers of the integrands in \eqref{appA1} are given by 
\begin{footnotesize}
\begin{align}\label{appA2}
\begin{split}
\hat{\theta}_{i,t}
:=
\begin{cases}
\Big(\ta_i \gamma(2 \kappa +\nu_4) (M \nu_0 (\overline{\kappa}-\bar{\nu}_3)+\overline{M} \bar{\nu}_0 (\kappa -\nu_3))+\taS\gamma \big(2 \kappa  \nu_0 (\bar{\nu}_3-\overline{\kappa})-\overline{\kappa} \nu_0 \nu_4\\
\quad +\nu_0 \nu_4 \bar{\nu}_3+\overline{M} (\kappa -\nu_3) (\nu_1 \bar{\nu}_0-\nu_0 \bar{\nu}_1)\big)+B_t \overline{M} (\kappa -\nu_3) (\nu_2 \bar{\nu}_0-\nu_0 (\bar{\nu}_2+2 \overline{\kappa} \epsilon ))\Big)\\
\quad \Big/\Big(2 (\kappa -\nu_3) ((M+1) \nu_0 (\overline{\kappa}-\bar{\nu}_3)+\overline{M} \bar{\nu}_0 (\kappa -\nu_3))\Big)
,\quad i\in\{1,...,M\},
\vspace{0.1cm}
\\
\frac{\taS\gamma \big(M \nu_0 \bar{\nu}_1- \bar{\nu}_0 (2 \kappa +M \nu_1+\nu_4)\big)+B_t M (\nu_0 \bar{\nu}_2-\nu_2 \bar{\nu}_0+2 \overline{\kappa} \nu_0 \epsilon )}{2 (M \nu_0 (\overline{\kappa}-\bar{\nu}_3)+(\overline{M}+1) \bar{\nu}_0 (\kappa -\nu_3))}
,\quad i\in\{M+1,...,M+\overline{M}\}.
\end{cases}
\end{split}
\end{align}
\end{footnotesize}Summing \eqref{appA2} over $i\in\{1,...,M+\overline{M}\}$ shows that the stock market clears in the sense that \eqref{def:clearing_eq} holds when
\begin{align}\label{nubar0}
\bar{\nu}_0&:= \frac{\nu_0(\overline{\kappa}-\bar{\nu}_3)}{\kappa-\nu_3}.
\end{align}

Definition \eqref{nubar0} allows us to re-write the second-order conditions in \eqref{appAA2} as
\begin{align}\label{appA3}
\begin{split}
\frac{(\kappa -\nu_3) (M+\overline{M}+1)}{M+\overline{M}-1}>0,\\
\frac{(\overline{\kappa}-\bar{\nu}_3) (M+\overline{M}+1)}{M+\overline{M}-1}>0.
\end{split}
\end{align}
Because we have assumed \eqref{SOC}, the second-order conditions \eqref{appA3} hold. Consequently, the pointwise optimizers are given in \eqref{appA2}.

To ensure that the resulting stock-price drift processes $\mu^{\hat{\theta}_i}_t$ are the same for all investors $i\in \{1,...,M+\overline{M}\}$ and, in particular, ensuring that the { individual }private targets $(\ta_1,...,\ta_M)$ do not appear in $\mu^{\hat{\theta}_i}_t$, we define the deterministic functions
\begin{align}\label{nu4}
\begin{split}
\nu_4&:= -\tfrac{2\kappa\big(2\kappa +(M+\overline{M}-2)(1+M+\overline{M})\nu_3\big)}{(M+\overline{M})\big(2(M+\overline{M})\kappa - (1+M+\overline{M})\nu_3\big)},\\
\bar{\nu}_1 &:= \tfrac{(\overline{\kappa}-\bar{\nu}_3)\nu_1}{\kappa-\nu_3} + \tfrac{(\overline{\kappa}-\bar{\nu}_3)( \nu_3-\bar{\nu}_3 - (M+\overline{M})(2\kappa-2\overline{\kappa}-\nu_3+\bar{\nu}_3))(2 \kappa +\nu_4) }{(\kappa-\nu_3)(M \overline{M}(2(\kappa+\overline{\kappa})-\nu_3)+\overline{M}(2\overline{M} \kappa - (1+\overline{M})\nu_3)+M^2 (2\overline{\kappa}-\bar{\nu}_3)-M(1+\overline{M})\bar{\nu}_3)},\\
\bar{\nu}_2 &:= \tfrac{(\overline{\kappa}-\bar{\nu}_3)\nu_2}{\kappa-\nu_3} + \tfrac{2\epsilon \overline{\kappa}(-2\overline{M}(M+\overline{M})(\kappa- \overline{\kappa})-2\overline{\kappa}+\overline{M}(1+M+\overline{M})\nu_3+2\bar{\nu}_3-((M-1)M+3M \overline{M}+2\overline{M}^2)\bar{\nu}_3)}{M \overline{M}(2(\kappa+\overline{\kappa})-\nu_3)+\overline{M}(2\overline{M} \kappa - (1+\overline{M})\nu_3)+M^2 (2\overline{\kappa}-\bar{\nu}_3)-M(1+\overline{M})\bar{\nu}_3}.
\end{split}
\end{align}
By inserting \eqref{nubar0}  and \eqref{nu4} into \eqref{appA2} and simplifying, gives
\begin{footnotesize}
\begin{align}\label{appA4}
\hat{\theta}_{i,t}
=
\begin{cases}
\frac{2 (M+\overline{M}-1)}M \Big(\frac{\overline{M} (\taS\gamma \kappa -B_t \overline{\kappa} M \epsilon )}{M^2 (2 \overline{\kappa}-\bar{\nu}_3)+M \overline{M} (2 (\kappa +\overline{\kappa})-\nu_3)-M (\overline{M}+1) \bar{\nu}_3+\overline{M} (2 \kappa  \overline{M}-\nu_3 (\overline{M}+1))}
\\
\quad+\frac{\gamma \kappa  (\ta_i M-\taS)}{2 \kappa  (M+\overline{M})-\nu_3 (M+\overline{M}+1)}\Big),\quad i\in\{1,...,M\},
\vspace{0.1cm}
\\
\frac{2 (M+\overline{M}-1) (B_t \overline{\kappa} M \epsilon -\taS\gamma \kappa )}{M^2 (2 \overline{\kappa}-\bar{\nu}_3)+M \overline{M} (2 (\kappa +\overline{\kappa})-\nu_3)-M (\overline{M}+1) \bar{\nu}_3+\overline{M} (2 \kappa  \overline{M}-\nu_3 (\overline{M}+1))}
,\quad i\in\{M+1,...,M+\overline{M}\}.
\end{cases}
\end{align}
\end{footnotesize}Because the remaining functions $\nu_0, \nu_1,$ and $\nu_2$ do not appear in \eqref{appA4}, these functions are irrelevant in the sense that different $\nu_0, \nu_1,$ and $\nu_2$ functions all produce  the same equilibrium prices and investor holdings provided $\nu_0\neq0$. 
The functions $\alpha_1, \alpha_2,\alpha_3, \bar{\alpha}_1,\bar{\alpha}_2$ in \eqref{optimaltheta_alphas} are found by matching $(\taS, B_t, \ta_i)$ coefficients in \eqref{appA4}, which also produces the expression for $\hat{\theta}_{i,t}$ in \eqref{optimaltheta_ismall}. 

At this point, the stock-price drift processes $\mu^{\hat{\theta}_i}_t$ are all identical, and we define $\hat{\mu}_t$ as their common value. The representation \eqref{dSmuuuu} of $\hat{\mu}_t$ follows from inserting \eqref{appA4} into $\mu^{\hat{\theta}_i}_t$ and matching $(\taS,B_t)$ coefficients. The resulting functions $\mu_1(t)$ and $\mu_2(t)$ are given in \eqref{mu12}, which, we show next, are integrable. The bounds 
\eqref{SOCCCC} produce
\begin{align} \label{mu122222} 
\begin{split}
&|\mu_1|\le \min\Big\{\frac{2\kappa}{M}, \frac{2(2(M+\overline{M})\overline{\kappa}-(1+M+\overline{M})\bar{\nu}_3)}{\overline{M}}\Big\}, \\
&|\mu_2|\le \min \Big\{2\overline{\kappa}, \frac{2\overline{M} (2(M+\overline{M})\kappa-(1+M+\overline{M})\nu_3)}{M}\Big\}.
\end{split}
\end{align}
The integrability \eqref{square_integrable} assumed of $\kappa$ and $\overline{\kappa}$ ensure integrability of the upper bounds in \eqref{mu122222}.
Therefore, the ODEs \eqref{ODEs_equilibrium} can be solved by integrating. To see that the functions $\alpha_1, \alpha_2, \alpha_3, \bar{\alpha}_1, \bar{\alpha}_2$ in \eqref{optimaltheta_alphas} are uniformly bounded we again use \eqref{mu122222}:
\begin{align*}
&|\alpha_1|\le \frac{M+\overline{M}-1}{\kappa} |\mu_1|\le \frac{2(M+\overline{M}-1)}M, \\ 
&|\alpha_2|\le\frac{2\overline{M}(M+\overline{M}-1)}{M},\\
&|\alpha_3|\le2(M+\overline{M}-1),\\
&|\bar{\alpha}_1|\le\frac{2(M+\overline{M}-1)}{\overline{M}},\\
&|\bar{\alpha}_2|\le\frac{M(M+\overline{M}-1)}{\overline{M}\overline{\kappa}} |\mu_2|\le \frac{2M(M+\overline{M}-1)}{\overline{M}}.
\end{align*}
Because $\alpha_1, \alpha_2,\alpha_3, \bar{\alpha}_1,\bar{\alpha}_2$ are bounded functions, the process $\hat{\theta}_{i,t}$  in \eqref{optimaltheta_ismall} is admissible in the sense of Definition \ref{def_admis}, and its optimality follows.

The last step of the proof establishes the terminal price condition \eqref{terminalS}. It\^o's lemma produces the dynamics of $\hat{S}_t$ in \eqref{S_equilibrium} to be
\begin{align}
\begin{split}
d\hat{S}_t &=g'(t)\taS dt + dD_t +  \zeta'(t)\epsilon B_tdt + \zeta(t)\epsilon dB_t\\
&= \hat{\mu}_t dt + dD_t+  \zeta(t) \epsilon dB_t,
\end{split}
\end{align}
where the last equality uses \eqref{ODEs_equilibrium}. The terminal conditions in \eqref{ODEs_equilibrium} produce the terminal price restriction \eqref{terminalS}.

$\endproof$

\noindent\emph{Proof of Theorem \ref{welfaremax}:} The Cauchy-Schwartz inequality ensures that when $M\ge2$ and $(\ta_1,...,\ta_M)$ are not perfectly correlated, we have 
\begin{align}\label{CS01}
M\sum_{i=1}^M \E[\ta_i^2]>\E[\ta_\Sigma^2].
\end{align}
 %The following result is stated in the deterministic target setting of Theorem \ref{thm:Main} but its conclusions continue to hold in the stochastic target settings of Theorem \ref{thm:Main_BM}  and Theorem \ref{thm:Main_gamma}.
We can rewrite \eqref{welfaremax0} as
\begin{align} 
\sum_{i=1}^{M+\overline{M}} \E[\textrm{CE}_i] &= \int_0^1 F\big(\nu_3(t),\bar{\nu}_3(t)\big)dt, 
\end{align}
where the function $F: \mathbb{R}^2\mapsto \mathbb{R}$ is defined by
\begin{equation}
\begin{split}\label{F_def}
F\big(\nu_3(t),\bar{\nu}_3(t)\big):&=\sum_{i=1}^{M} \E\big[\hat{\theta}_{i,t}\hat{\mu}_t- \kappa(t)\big(\gamma(t) \ta_i - \hat{\theta}_{i,t}\big)^2\big]\\
&\quad + \sum_{i=M+1}^{M+\overline{M}} \E\big[ \hat{\theta}_{i,t}\hat{\mu}_t- \overline{\kappa}(t)(\hat{\theta}_{i,t}- \epsilon B_t)^2 \big]\\
&=-\sum_{i=1}^{M} \E\big[\kappa(t)\big(\gamma(t) \ta_i - \hat{\theta}_{i,t}\big)^2\big]-\sum_{i=M+1}^{M+\overline{M}}\E\big[\overline{\kappa}(t)(\hat{\theta}_{i,t}- \epsilon B_t)^2 \big],
\end{split}
\end{equation}
{ where $\hat{\theta}_{i,t}$ is as in Theorem \ref{thm:Main}. The equality in \eqref{F_def} follows from $\sum_{i=1}^{M+\bar{M}}\hat{\theta}_{i,t}=0$.}
 Therefore, our goal is to find the maximum point of $F(\nu_3,\bar{\nu}_3)$ under the restrictions  $\nu_3<\kappa$ and $\bar{\nu}_3<\overline{\kappa}$. We observe that
\begin{equation}\label{nu3bar_derivative}
\begin{split}
&\tfrac{\partial }{\partial \bar{\nu}_3}F(\nu_3,\bar{\nu}_3)\\&=-\tfrac{4\overline{M}(M+\overline{M}-1)(M+\overline{M}+1)(\E[\ta_\Sigma^2] \gamma^2 \kappa^2+M^2 t \epsilon^2 \overline{\kappa}^2)\big(M(1+M+\overline{M})\bar{\nu}_3+ \overline{M}(1+M+\overline{M})\nu_3-2\overline{M}\kappa-2M\overline{\kappa}\big)
}{\Big(\overline{M}\big(2(M+\overline{M})\kappa-(1+M+\overline{M})\nu_3\big)+M\big(2(M+\overline{M})\overline{\kappa}-(1+M+\overline{M})\bar{\nu}_3\big)\Big)^3}.
\end{split}
\end{equation}
%KL: I get (A.19)
{ Because the denominator in \eqref{nu3bar_derivative} is positive by \eqref{SOCCCC}, we have 
\begin{align}
\frac{\partial F}{\partial \bar{\nu}_3}(\nu_3,\bar{\nu}_3)>0,
\end{align}
for $(\nu_3,\bar{\nu}_3)$ satisfying}
\begin{align}
 \nu_3< \frac{2\kappa}{1+M+\overline{M}}-\frac{M(M+\overline{M}-1)\overline{\kappa}}{\overline{M}(1+M+\overline{M})}  \quad \textrm{and} \quad \bar{\nu}_3<\overline{\kappa}.
\end{align}
Therefore, 
\begin{align}\label{nu3cond1}
\sup_{\bar{\nu}_3<\overline{\kappa}}F(\nu_3,\bar{\nu}_3)=F(\nu_3, \overline{\kappa}) \quad \textrm{when} \quad    \nu_3< \frac{2\kappa}{1+M+\overline{M}}-\frac{M(M+\overline{M}-1)\overline{\kappa}}{\overline{M}(1+M+\overline{M})}.
\end{align}
Long, but elementary computations produce 
\begin{align}\label{nu3cond2}
\frac{\partial}{\partial \nu_3}F(\nu_3,\overline{\kappa})>0, \quad \textrm{for} \quad \nu_3< \frac{2\kappa}{1+M+\overline{M}}-\frac{M(M+\overline{M}-1)\overline{\kappa}}{\overline{M}(1+M+\overline{M})}. 
\end{align}
From \eqref{nu3cond1} and \eqref{nu3cond2}, we conclude that the maximum point of $F$  satisfies the inequality $\nu_3\geq \frac{2\kappa}{1+M+\overline{M}}-\frac{M(M+\overline{M}-1)\overline{\kappa}}{\overline{M}(1+M+\overline{M})}$, which is equivalent to 
\begin{align}\label{nu3cond3}
\frac{2\overline{M}\kappa+2M\overline{\kappa}-\overline{M}(1+M+\overline{M})\nu_3}{M(1+M+\overline{M})}\leq \overline{\kappa}.
\end{align}
Under the restriction in \eqref{nu3cond3}, the derivative \eqref{nu3bar_derivative} implies that we have
\begin{align}
\sup_{\bar{\nu}_3<\overline{\kappa}}F(\nu_3,\bar{\nu}_3)=F(\nu_3, \frac{2\overline{M}\kappa+2M\overline{\kappa}-\overline{M}(1+M+\overline{M})\nu_3}{M(1+M+\overline{M})}).
\end{align}

{ Taking the derivative with respect to $\nu_3$ of $F$ in \eqref{F_def} produces}
\begin{align}
\begin{split}
&\frac{\partial}{\partial \nu_3}F(\nu_3, \tfrac{2\overline{M}\kappa+2M\overline{\kappa}-\overline{M}(1+M+\overline{M})\nu_3}{M(1+M+\overline{M})}) \\
&=-\tfrac{4(M\sum_{i=1}^M \E[\ta_i^2]-\E[\ta_\Sigma^2])(M+\overline{M}-1)(1+M+\overline{M})\gamma^2 \kappa^2\big((1+M+\overline{M})\nu_3-2\kappa\big)}{M\big(2(M+\overline{M})\kappa-(1+M+\overline{M})\nu_3\big)^3}.
\end{split}
\end{align}
{ Because of \eqref{CS01}, we can }conclude that $F(\nu_3, \tfrac{2\overline{M}\kappa+2M\overline{\kappa}-\overline{M}(1+M+\overline{M})\nu_3}{M(1+M+\overline{M})})$ is maximized at $\nu_3^*=\frac{2 \kappa}{1 + M + \overline{M}}$, and the corresponding $\bar{\nu}_3^*$ is $\frac{2 \overline{\kappa}}{1 + M + \overline{M}}$. This maximizer is unique by the previous arguments. 

Finally, inserting $(\nu_3^*,\bar{\nu}_3^*)$ into $F$ in \eqref{F_def} produces the total welfare
\begin{align}\label{social_welfare_max}
\sup_{\nu_3(t),\bar{\nu}_3(t)}\sum_{i=1}^{M+\overline{M}}\E[\text{CE}_i]=-\int_0^1 \frac{\kappa(t)\overline{\kappa}(t)\big( \gamma(t)^2\E[\ta_\Sigma^2] +\epsilon^2 \overline{M}^2 t\big) }{\overline{M} \kappa(t)+M \overline{\kappa}(t)} dt,
\end{align}
the optimal holding strategies in \eqref{social_welfare_thetas}, and the common drift in \eqref{common_drift_Radner}.  

$\endproof$

\noindent \emph{Proof of Theorem \ref{thm:Main_BM}:} This proof is similar to the proof of Theorem \ref{thm:Main}, so here we only outline the key difference. It\^o's product rule produces the dynamics of the right-hand-side of \eqref{S_VWAPBM} to be
\begin{align}\label{eq:EtDTgamma1}
\begin{split}
& h'(t)\taS dt + dD_t + \zeta'(t)\epsilon B_tdt+ \zeta(t)\epsilon dB_t + \taS \big(\sigma(t) d\gamma_t + \gamma_t\sigma'(t)dt\big) \\
&= 
\Big( h'(t) \taS+\zeta'(t)\epsilon B_t+\taS \big(\sigma(t) \frac{1-\gamma_t}{1-t} + \sigma'(t)\gamma_t\big)\Big)dt+dD_t + \zeta(t)\epsilon dB_t+\sigma(t)\taS dZ_t\\
&= 
\hat{\mu}_tdt+dD_t +\zeta(t)\epsilon  dB_t+\sigma(t)\taS dZ_t,
\end{split}
\end{align}
where the last equality uses the ODEs \eqref{ODEs_equilibrium_BM}. The terminal conditions in \eqref{ODEs_equilibrium_BM} produce the terminal price restriction \eqref{terminalS}.

By computing the derivative in formula \eqref{sigmaSgamma}, we see that the ODE for $\sigma$ in \eqref{ODEs_equilibrium_BM} holds. The zero terminal condition for $\sigma$ follows from
$$
|\sigma(t)| \le \frac1{1-t} \int_t^1 (1-u) |\mu_1(u)|du \le \int_t^1 |\mu_1(u)|du,\quad t\in[0,1),
$$
which converges to zero as $t\uparrow 1$ because $\mu_1(u)$ is integrable over $u\in[0,1]$. To see that $\frac{\sigma(t)}{1-t}$ is integrable, the representation \eqref{sigmaSgamma} produces
\begin{align}
\begin{split}
\int_0^1 |\frac{\sigma(t)}{1-t}| dt &\le \int_0^1 \int_t^1\frac{1-u}{(1-t)^2}|\mu_1(u)|du dt\\
&= \int_0^1 u|\mu_1(u)|du<\infty,
\end{split}
\end{align}
where the equality uses Tonelli's theorem. 

$\endproof$

\noindent \emph{Proof of Theorem \ref{thm:Main_gamma}:} This proof is similar to the proofs of Theorems \ref{thm:Main} and  \ref{thm:Main_BM}, so here we only outline the key difference. It\^o's product rule produces the dynamics of the right-hand-side of \eqref{S_VWAPBM} to be
\begin{align}\label{eq:EtDTgamma2}
\begin{split}
& \taS h'(t)dt + dD_t +\zeta'(t) \epsilon B_tdt+\zeta(t)\epsilon dB_t + \taS \big(\sigma(t) d\gamma_t + \sigma'(t)\gamma_{t-}dt\big) \\
&= 
\Big( h'(t)\taS+\zeta'(t)\epsilon B_t+\taS \big(\sigma(t) \tfrac{1-\gamma_{t-}}{1-t} + \sigma'(t)\gamma_{t-}\big)\Big)dt\\
&+dD_t + \zeta(t)\epsilon dB_t+\sigma(t)\taS\big(d\gamma_t-\tfrac{1-\gamma_{t-}}{1-t}dt\big) \\
&= 
\hat{\mu}_tdt+dD_t + \zeta(t)\epsilon dB_t+\sigma(t)\taS\big(d\gamma_t-\tfrac{1-\gamma_{t-}}{1-t}dt\big),
\end{split}
\end{align}
where the last equality uses the ODEs \eqref{ODEs_equilibrium_BM}. The terminal conditions in \eqref{ODEs_equilibrium_BM} produce the terminal price restriction \eqref{terminalS}.

$\endproof$

\end{document}